%% file: Fear_PBI.tex
\documentclass[letterpaper,twocolumn,10pt]{article}\usepackage[]{graphicx}\usepackage[]{color}
\usepackage{usenix}

\usepackage{epsfig,endnotes,url}

\usepackage{xspace}

\usepackage[letterpaper]{geometry}
\usepackage[HideComments]{mycomment}

\usepackage[english]{babel}
\usepackage[latin1]{inputenc}
\usepackage{amsmath, amssymb, latexsym}

\usepackage{graphicx}
\usepackage{multirow}
\usepackage{booktabs}

\usepackage[caption=false]{subfig}

\usepackage{paralist}

\usepackage{balance}

\usepackage{rotating}

\usepackage{versions}
\includeversion{DocumentVersionTR}

\newtheorem{researchquestion}{RQ}

\newcommand{\const}[1]{\ensuremath{\mathsf{#1}\xspace}}
\newcommand{\vari}[1]{\ensuremath{\mathit{#1}\xspace}}

\newcommand{\CASCAde}{ERC Starting Grant CASCAde (GA n\textsuperscript{o}716980)}
\IfFileExists{upquote.sty}{\usepackage{upquote}}{}
\begin{document}



\newcommand{\descriptivesOffline}{
\begin{table}[!h]

\caption{\label{tab:desc_fear}\label{tab:descriptivesOffline}Descriptives of the lab experiment ($N_{L}=60$)}
\centering
\begin{tabular}[t]{lcccccc}
\toprule
\multicolumn{1}{c}{ } & \multicolumn{3}{c}{Condition Fear} & \multicolumn{3}{c}{Condition Happiness} \\
\cmidrule(l{3pt}r{3pt}){2-4} \cmidrule(l{3pt}r{3pt}){5-7}
  & PX Fear & PX Joviality & PBI & PX Fear & PX Joviality & PBI\\
\midrule
m & 1.55 & 2.71 & 3.94 & 1.13 & 3.27 & 4.01\\
sd & 0.84 & 1.23 & 0.93 & 0.26 & 1.18 & 0.89\\
\bottomrule
\end{tabular}
\end{table}
}
\newcommand{\descriptivesOnline}{
\begin{table}[!h]

\caption{\label{tab:desc_fear}\label{tab:descriptivesOnline}Descriptives of the MTurk experiment ($N_{M}=39$)}
\centering
\begin{tabular}[t]{lcccccc}
\toprule
\multicolumn{1}{c}{ } & \multicolumn{3}{c}{Condition Fear} & \multicolumn{3}{c}{Condition Happiness} \\
\cmidrule(l{3pt}r{3pt}){2-4} \cmidrule(l{3pt}r{3pt}){5-7}
  & PX Fear & PX Joviality & PBI & PX Fear & PX Joviality & PBI\\
\midrule
m & 1.43 & 2.37 & 4.32 & 1.53 & 2.23 & 4.31\\
sd & 0.80 & 1.16 & 1.08 & 0.78 & 1.28 & 1.10\\
\bottomrule
\end{tabular}
\end{table}
}

\newcommand{\descriptivesProlific}{
\begin{table}[!h]

\caption{\label{tab:desc_fear}\label{tab:descriptivesProlific}Descriptives of the Prolific experiment ($N_{M}=226$)}
\centering
\begin{tabular}[t]{lcccccc}
\toprule
\multicolumn{1}{c}{ } & \multicolumn{3}{c}{Condition Fear} & \multicolumn{3}{c}{Condition Happiness} \\
\cmidrule(l{3pt}r{3pt}){2-4} \cmidrule(l{3pt}r{3pt}){5-7}
  & PX Fear & PX Joviality & PBI & PX Fear & PX Joviality & PBI\\
\midrule
m & 1.59 & 2.09 & 4.23 & 1.25 & 2.40 & 4.20\\
sd & 0.75 & 1.02 & 0.85 & 0.46 & 1.07 & 0.86\\
\bottomrule
\end{tabular}
\end{table}
}


\newcommand{\plotTrigraphMC}{
\begin{figure*}[tb]
\subfloat[Joviality]{
\begin{minipage}{0.49\linewidth}
\processifversion{DocumentVersionSTAST}{\vspace{-2cm}}

\includegraphics[width=\linewidth]{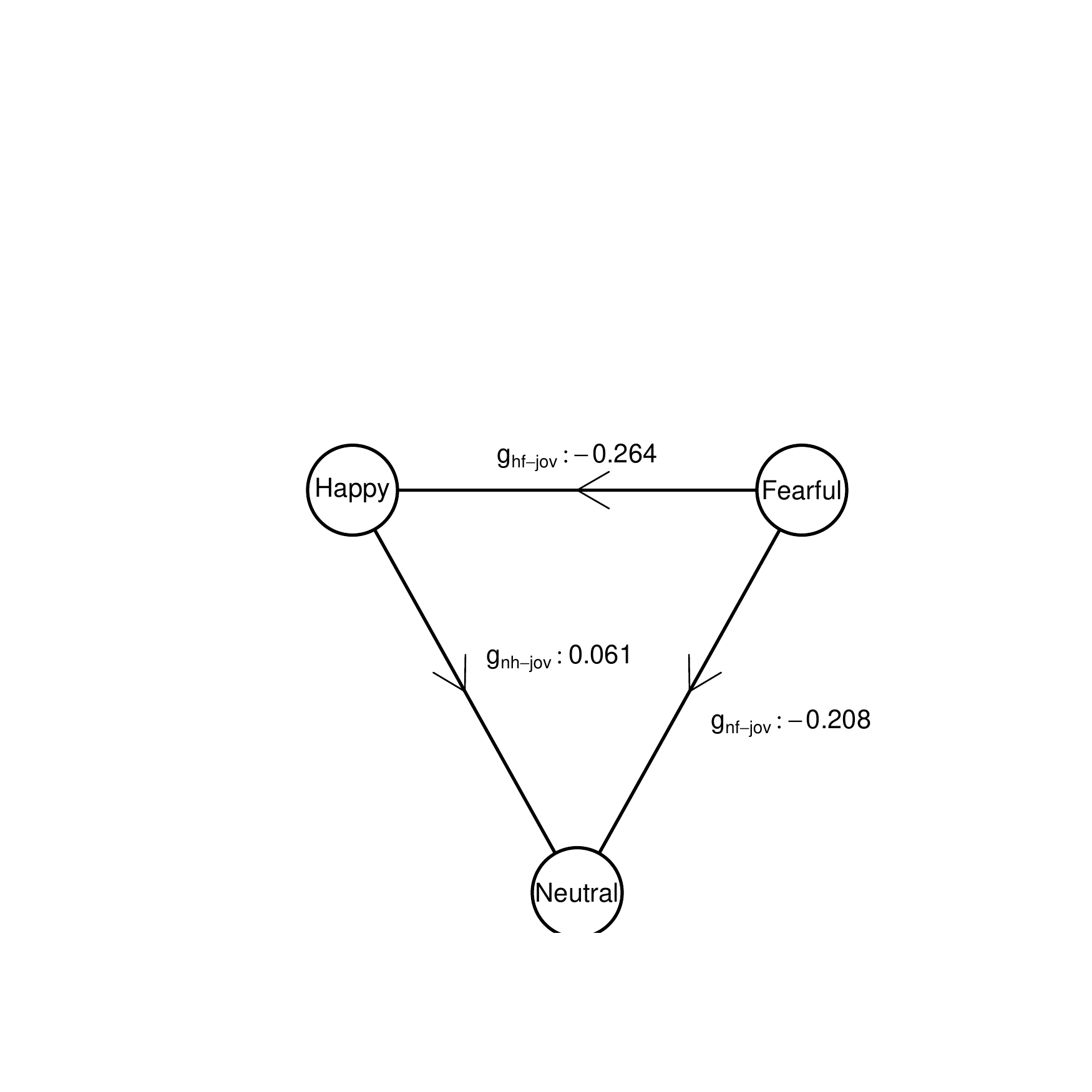} 
\processifversion{DocumentVersionSTAST}{\vspace{-1cm}}
\end{minipage}
}~
\subfloat[Fear]{
\begin{minipage}{0.49\linewidth}
\processifversion{DocumentVersionSTAST}{\vspace{-2cm}}

\includegraphics[width=\linewidth]{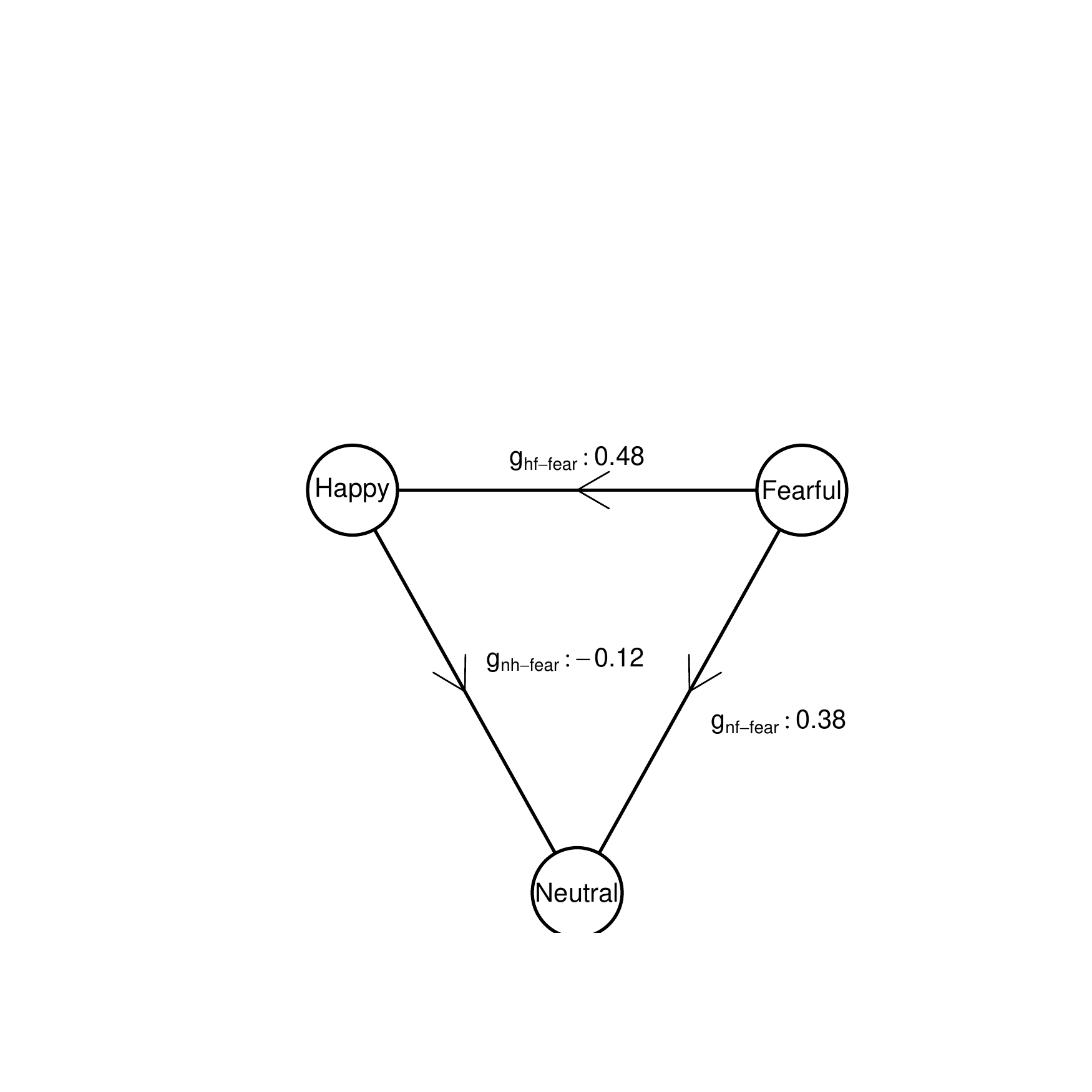} 
\processifversion{DocumentVersionSTAST}{\vspace{-1cm}}
\end{minipage}
}
\caption{Comparison of Hedges $g$ of manipulation checks wrt. joviality and fear.}\label{fig:plotTrigraphMC}
\end{figure*}
}


\newcommand{\plotTrigraphPBI}{
\begin{figure*}[tb]
\subfloat[PBI]{
\begin{minipage}{0.49\linewidth}
\processifversion{DocumentVersionSTAST}{\vspace{-2cm}}

\includegraphics[width=\linewidth]{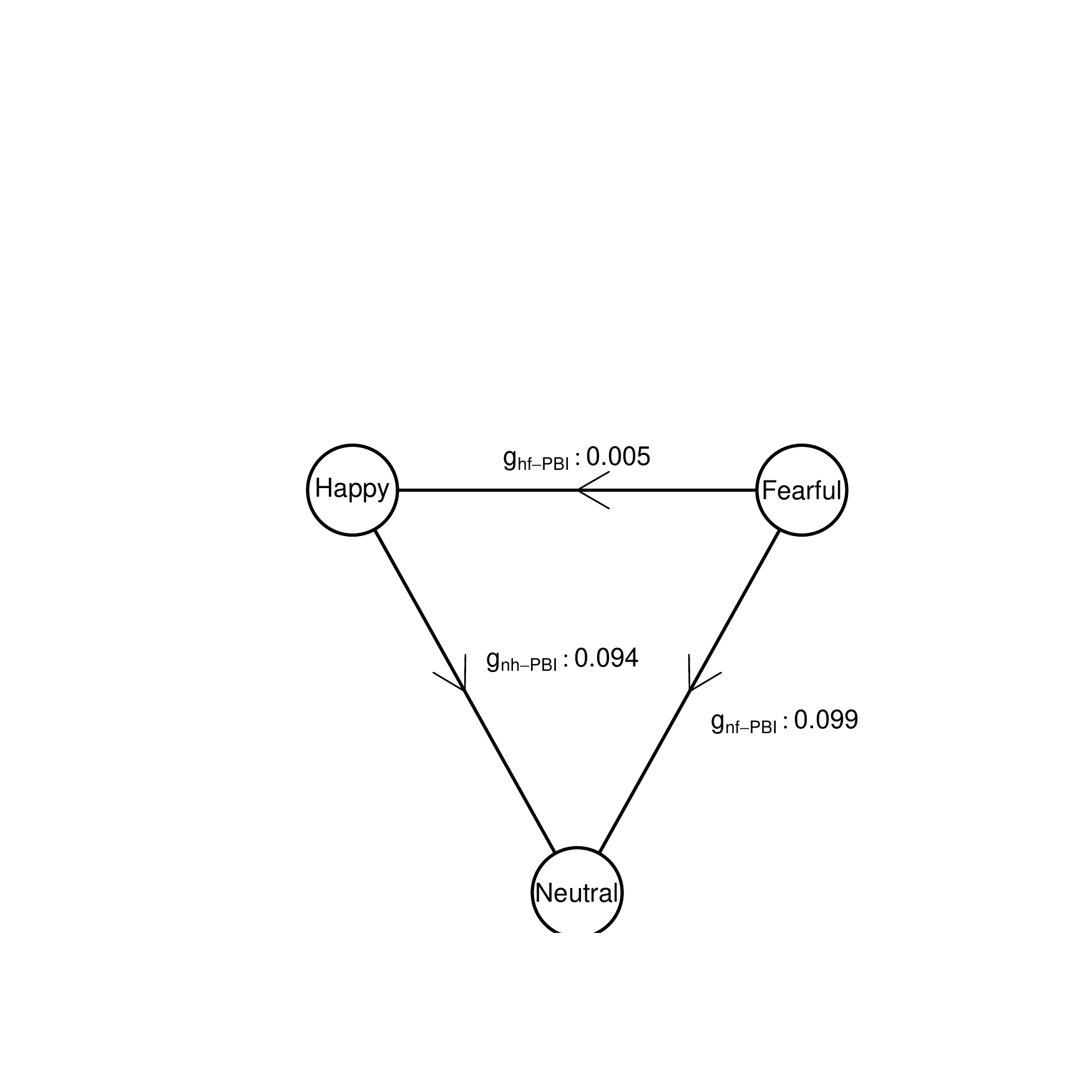} 
\processifversion{DocumentVersionSTAST}{\vspace{-1cm}}
\end{minipage}
}~
\subfloat[IDI]{
\begin{minipage}{0.49\linewidth}
\processifversion{DocumentVersionSTAST}{\vspace{-2cm}}

\includegraphics[width=\linewidth]{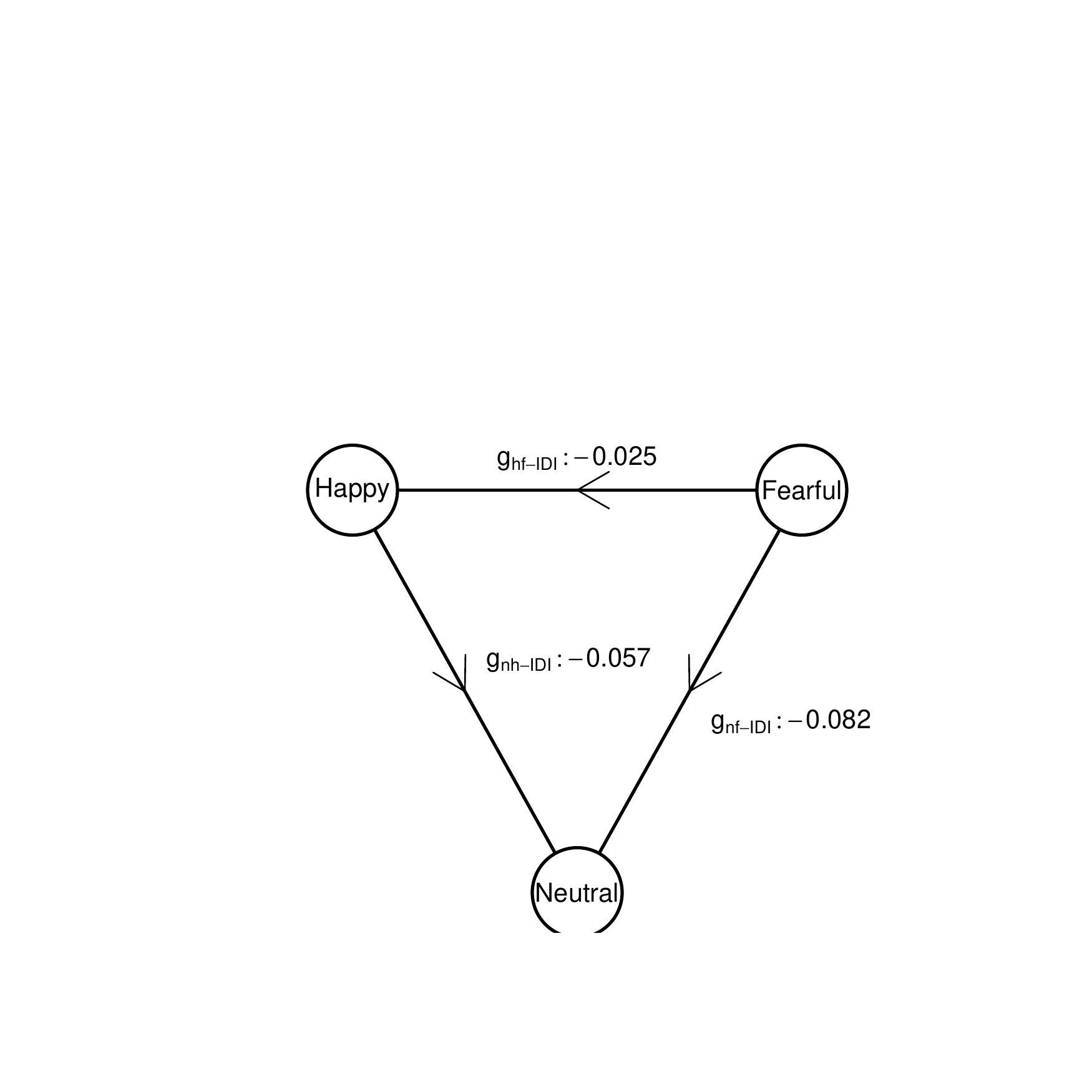} 
\processifversion{DocumentVersionSTAST}{\vspace{-1cm}}
\end{minipage}
}

\processifversion{DocumentVersionSTAST}{\vspace{-1.5cm}}
\subfloat[PI]{
\begin{minipage}{0.49\linewidth}
\processifversion{DocumentVersionSTAST}{\vspace{-1cm}}

\includegraphics[width=\linewidth]{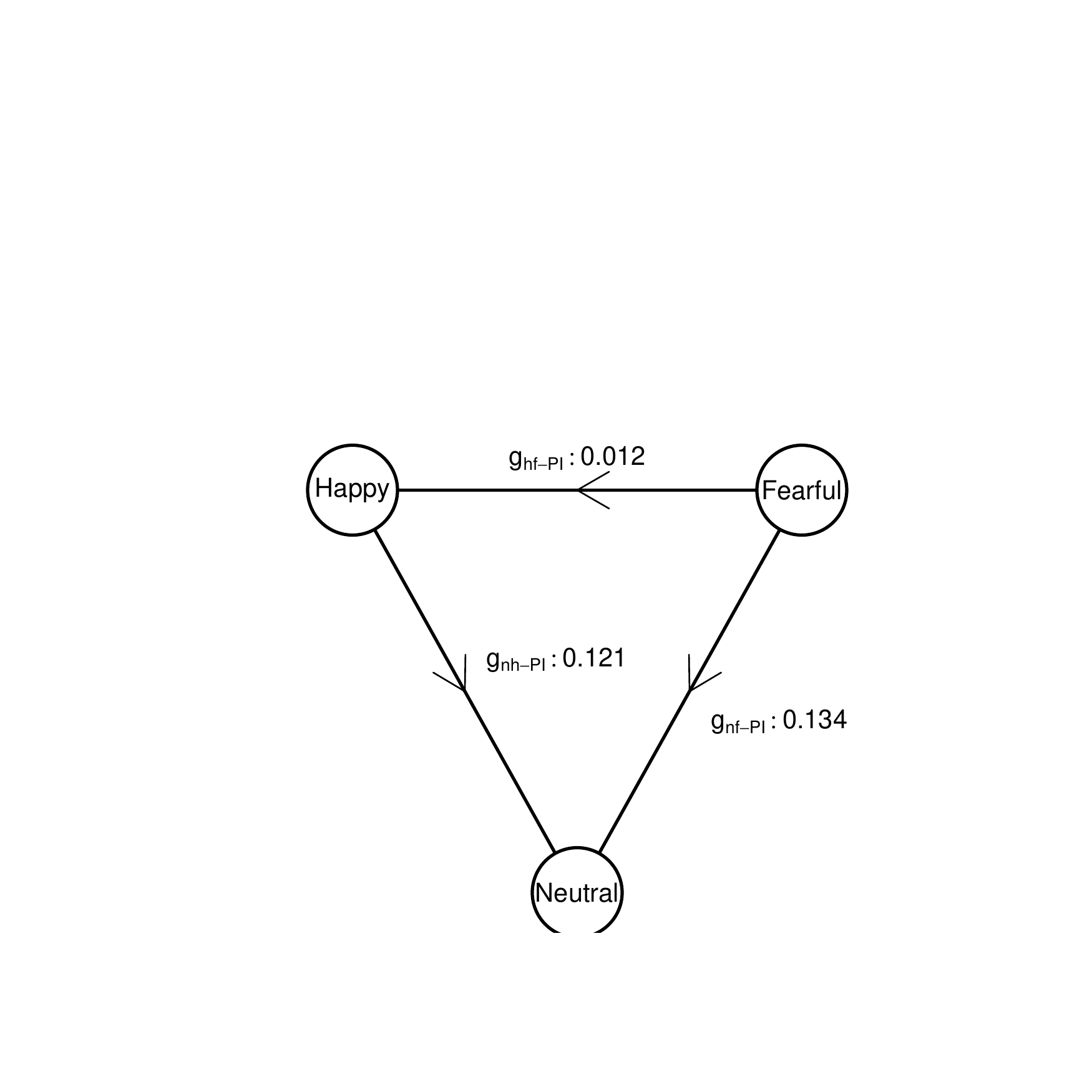} 
\processifversion{DocumentVersionSTAST}{\vspace{-1cm}}
\end{minipage}
}~
\subfloat[TI]{
\begin{minipage}{0.49\linewidth}
\processifversion{DocumentVersionSTAST}{\vspace{-1cm}}

\includegraphics[width=\linewidth]{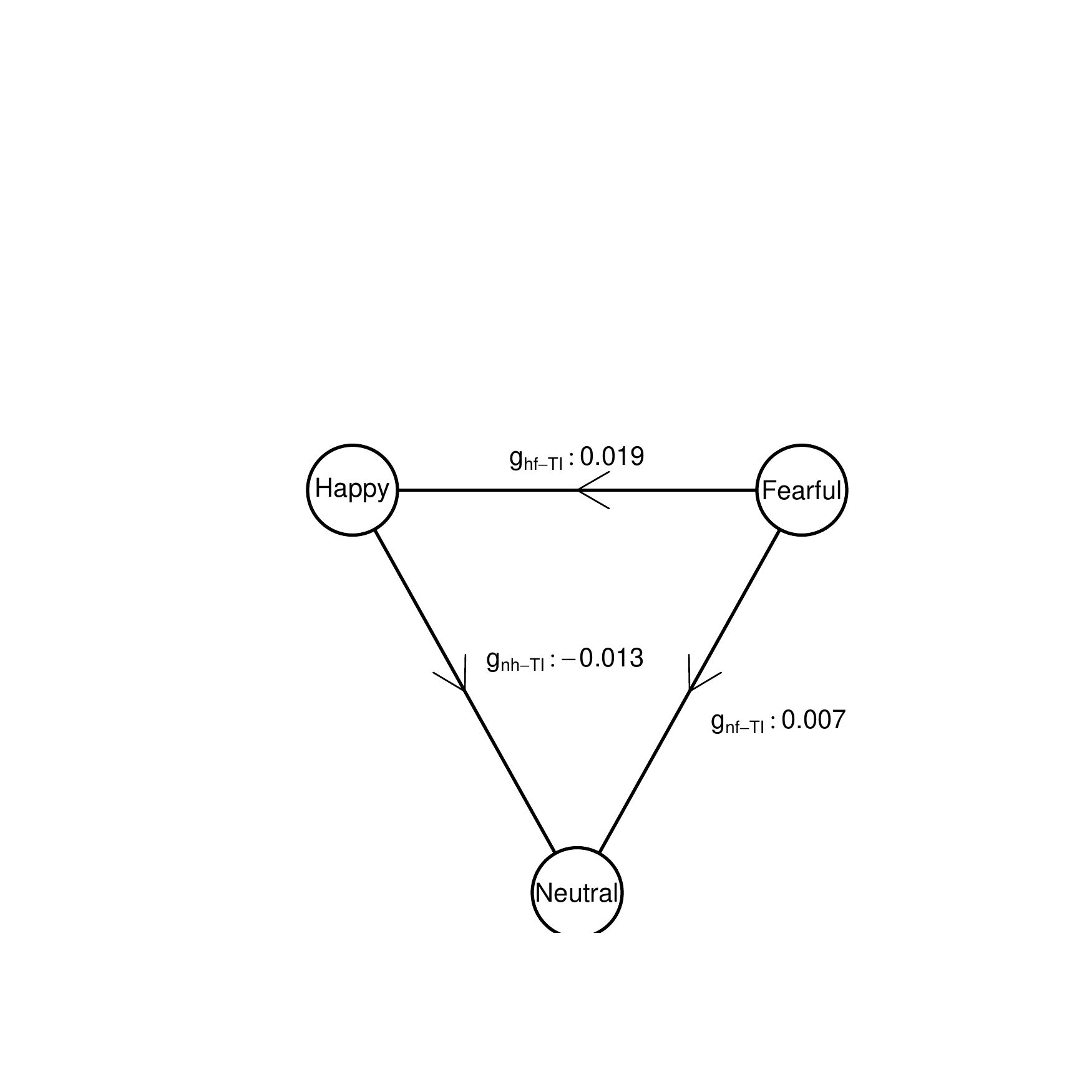} 
\processifversion{DocumentVersionSTAST}{\vspace{-1cm}}
\end{minipage}
}
\caption{Comparison of Hedges $g$ of PBI and its sub-scales Information Disclosure Intention (IDI), Protection Intention (PI), and Transaction Intention (TI).}\label{fig:plotTrigraphPBI}
\end{figure*}
}

\newcommand{\esAffPBI}{
\begin{figure}[htb]
\vspace{-2.5cm}
\includegraphics[width=\linewidth]{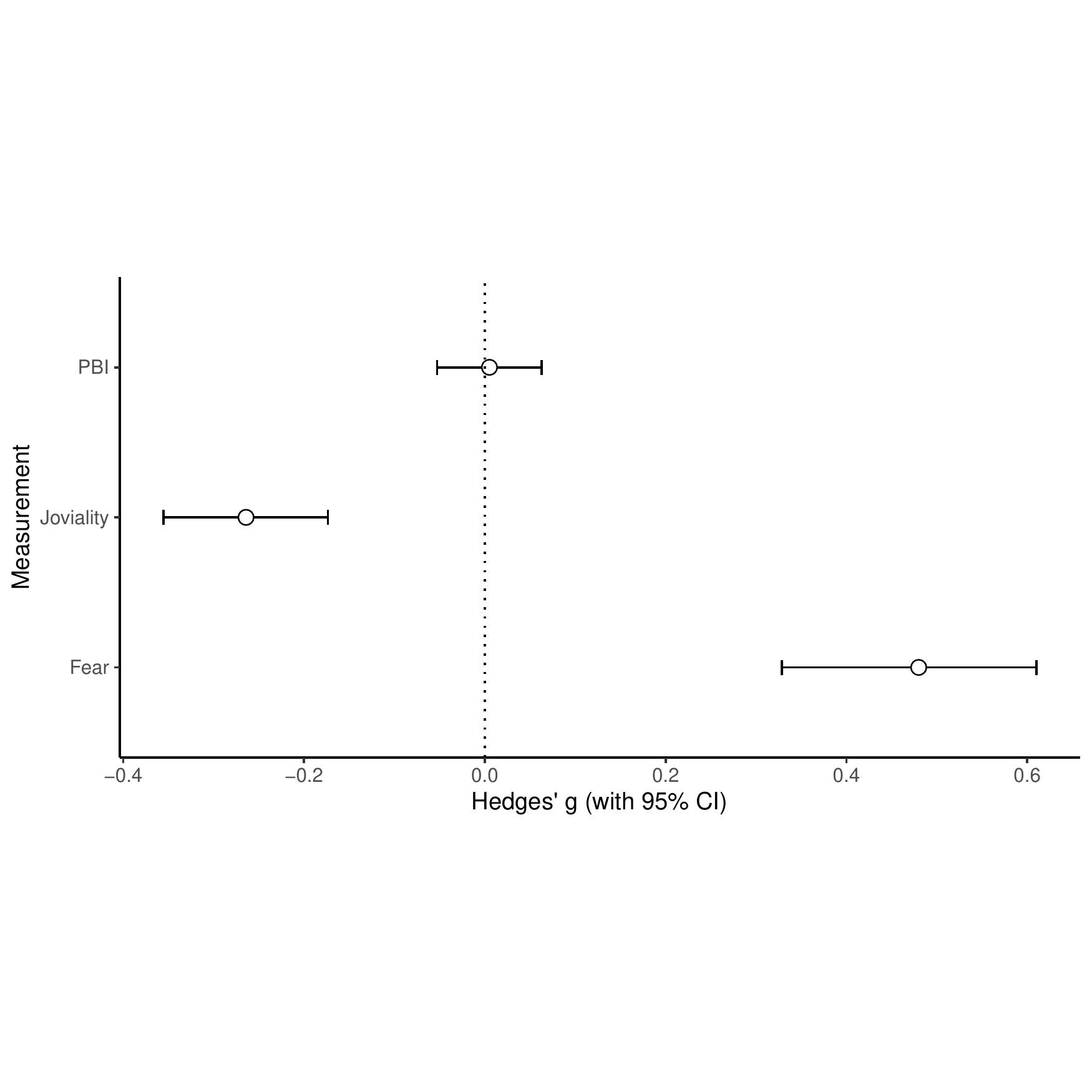} 
\vspace{-2.5cm}\caption{Comparison of the effects of the manipulations and of affects on PBI. (Hedges' $g_\const{av}$, 95\% Confidence Intervals).}
\label{fig:esAffPBI}
\end{figure}
}

\newcommand{\corrgramAffPbiOffline}{
\begin{figure*}[htb]

\includegraphics[width=\linewidth]{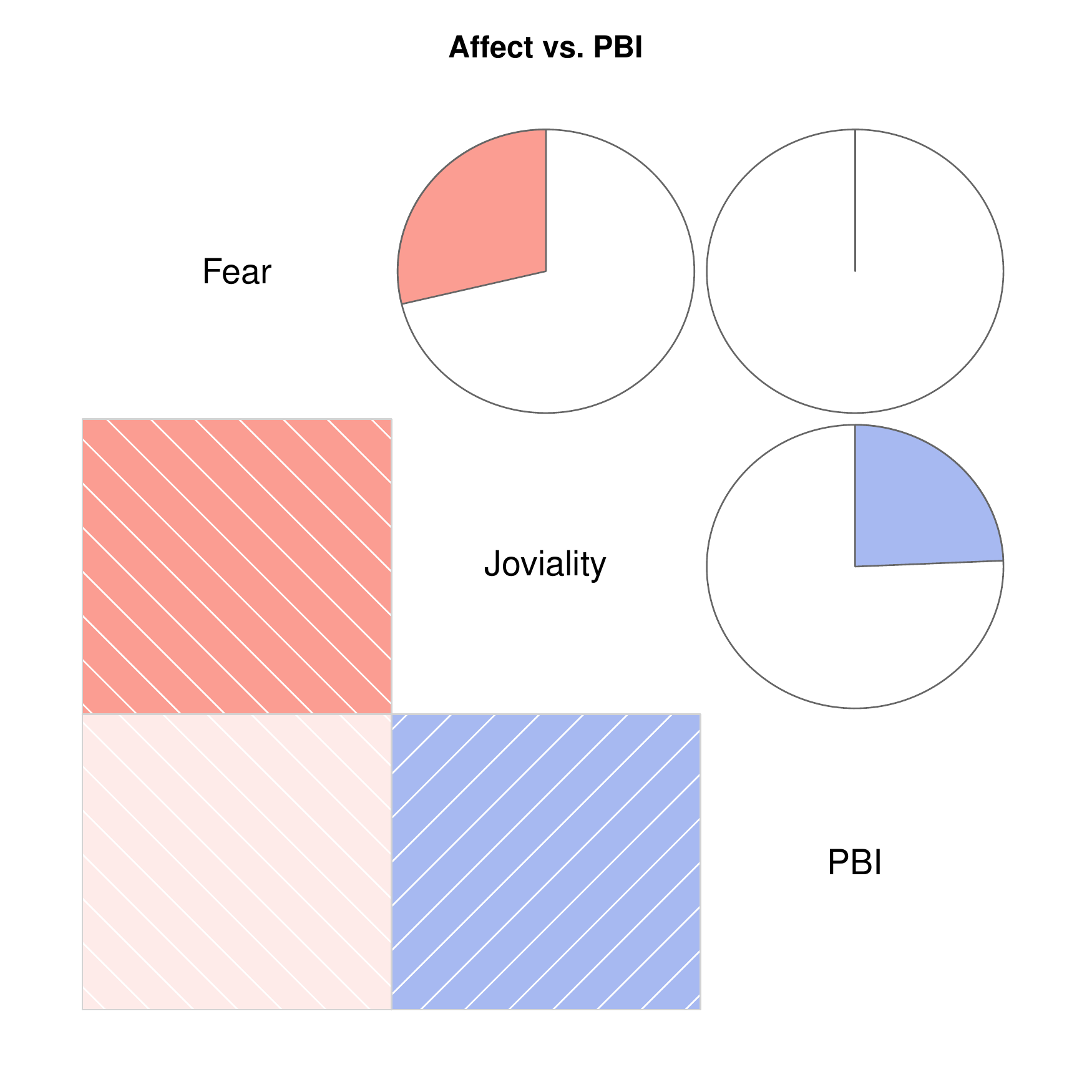} 
\caption{Corrgram of the variables irrespective of conditions from the Lab dataset.}
\label{fig:corrgramAffPBOffline}
\end{figure*}
}

\newcommand{\corrgramAffPbiOnline}{
\begin{figure*}[htb]

\includegraphics[width=\linewidth]{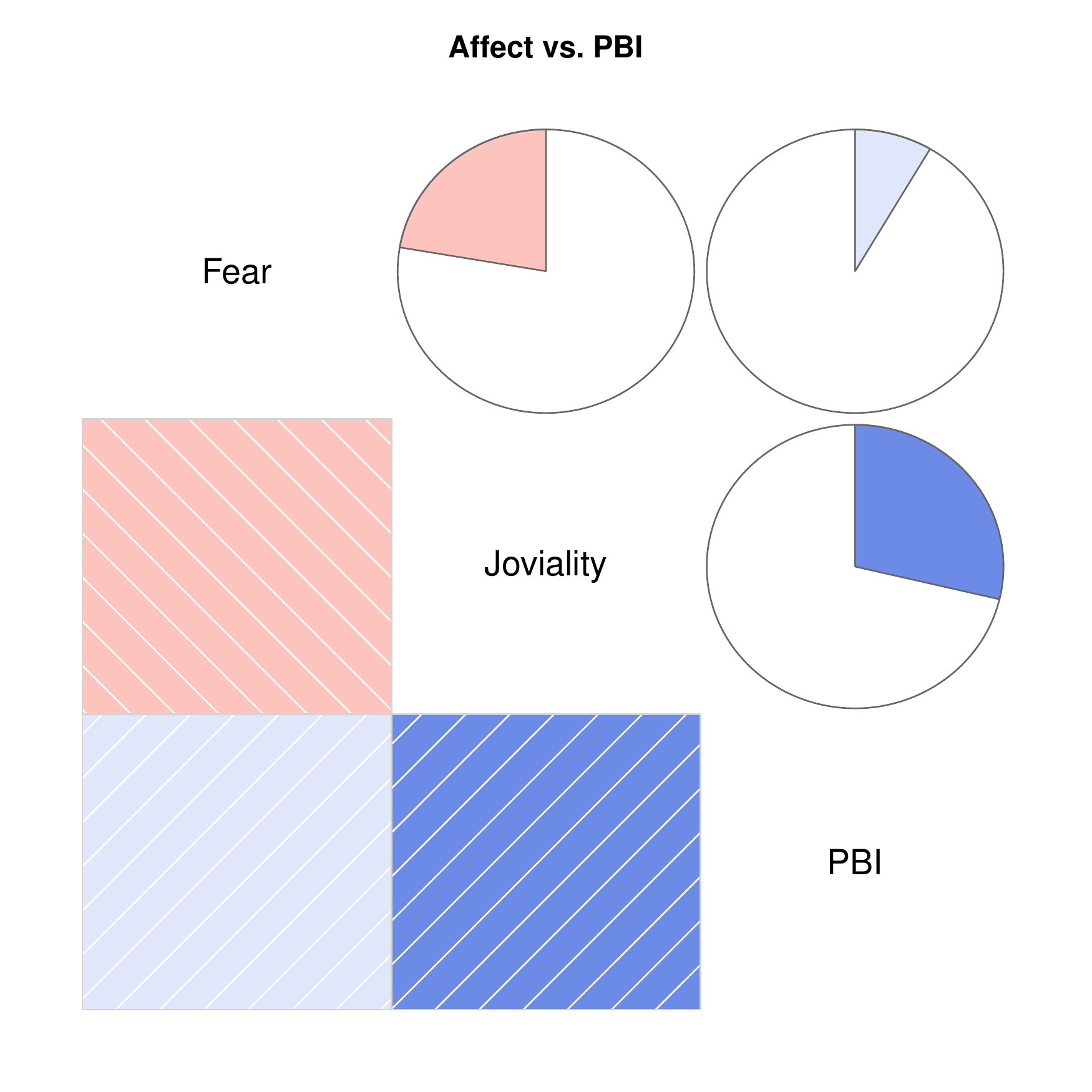} 
\caption{Corrgram of the variables irrespective of conditions from the MTurk dataset.}
\label{fig:corrgramAffPBOnline}
\end{figure*}
}

\newcommand{\interactConditionFear}{
\begin{figure*}[htb]

\includegraphics[width=\linewidth]{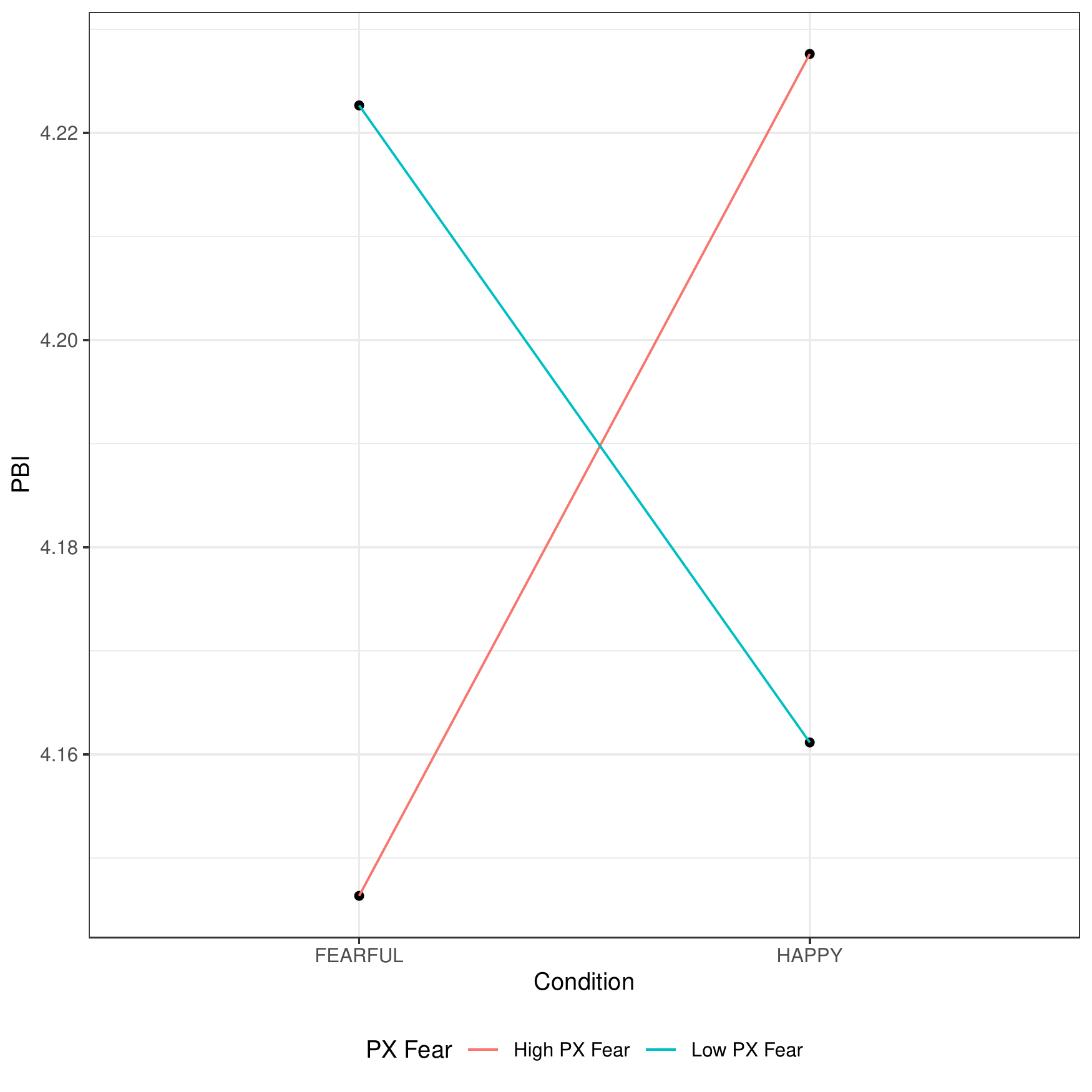} 
\caption{Interaction plot on PX Fear and Condition.}
\label{fig:interactConditionFear}
\end{figure*}
}
\newcommand{\interactsPBI}{
\begin{figure*}[htb]
\centering
\subfloat[PX Fear]{
\begin{minipage}{0.33\linewidth}
\centering
\includegraphics[width=\linewidth]{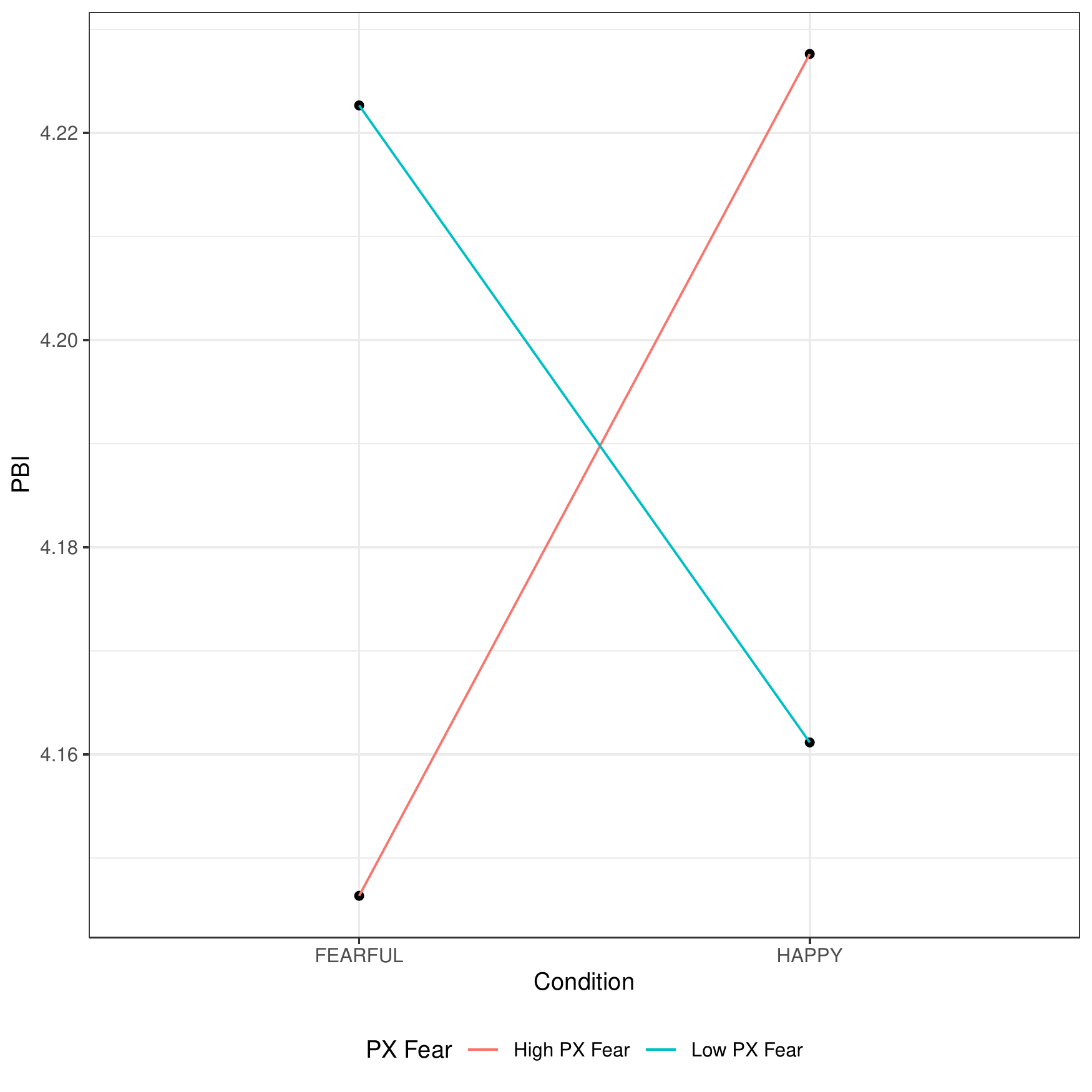} 
\end{minipage}
\label{fig:interactPBIFear}
}~
\subfloat[PX Joviality]{
\begin{minipage}{0.33\linewidth}
\centering
\includegraphics[width=\linewidth]{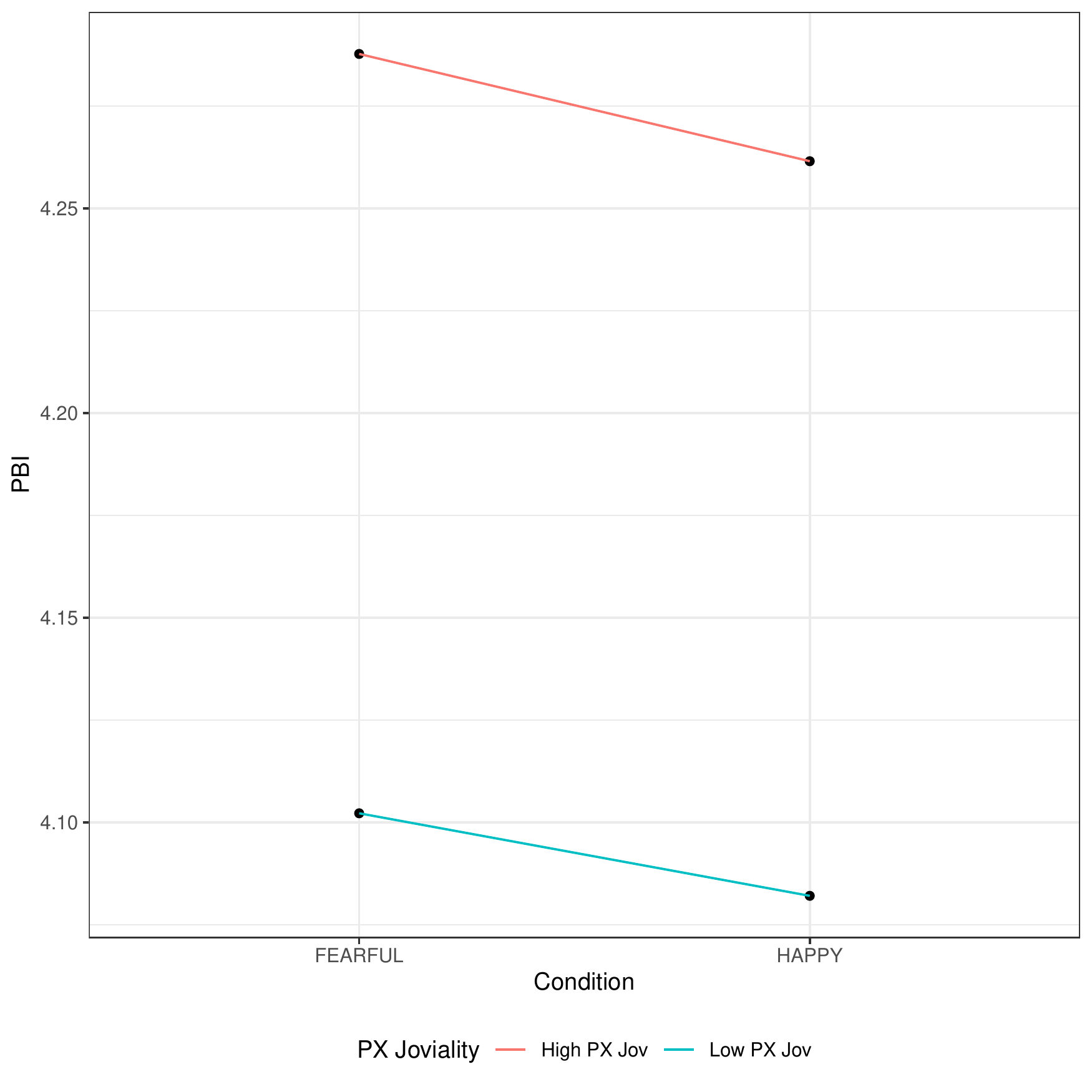} 
\end{minipage}
\label{fig:interactPBIJoviality}
}

\caption{Interaction plots of PBI on PX Affects and Condition.}
\label{fig:interactsPBI}
\end{figure*}
}
\newcommand{\interactsOverall}{
\begin{figure*}[htb]
\centering
\subfloat[IDI/Fear]{
\begin{minipage}{0.3\textwidth}

\includegraphics[width=\linewidth]{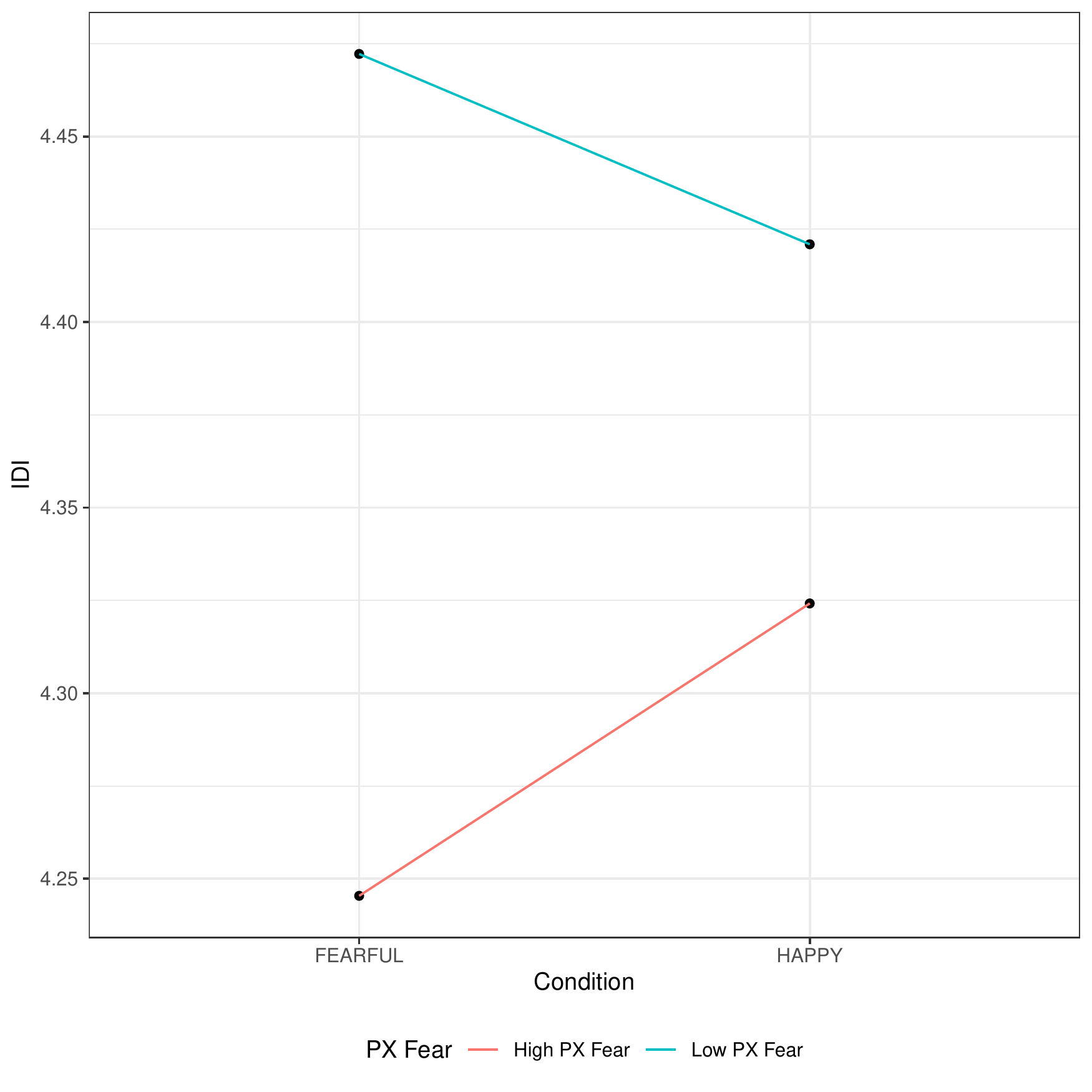} 
\label{fig:interactIDIFear}
\end{minipage}
}~
\subfloat[PI/Fear]{
\begin{minipage}{0.3\textwidth}

\includegraphics[width=\linewidth]{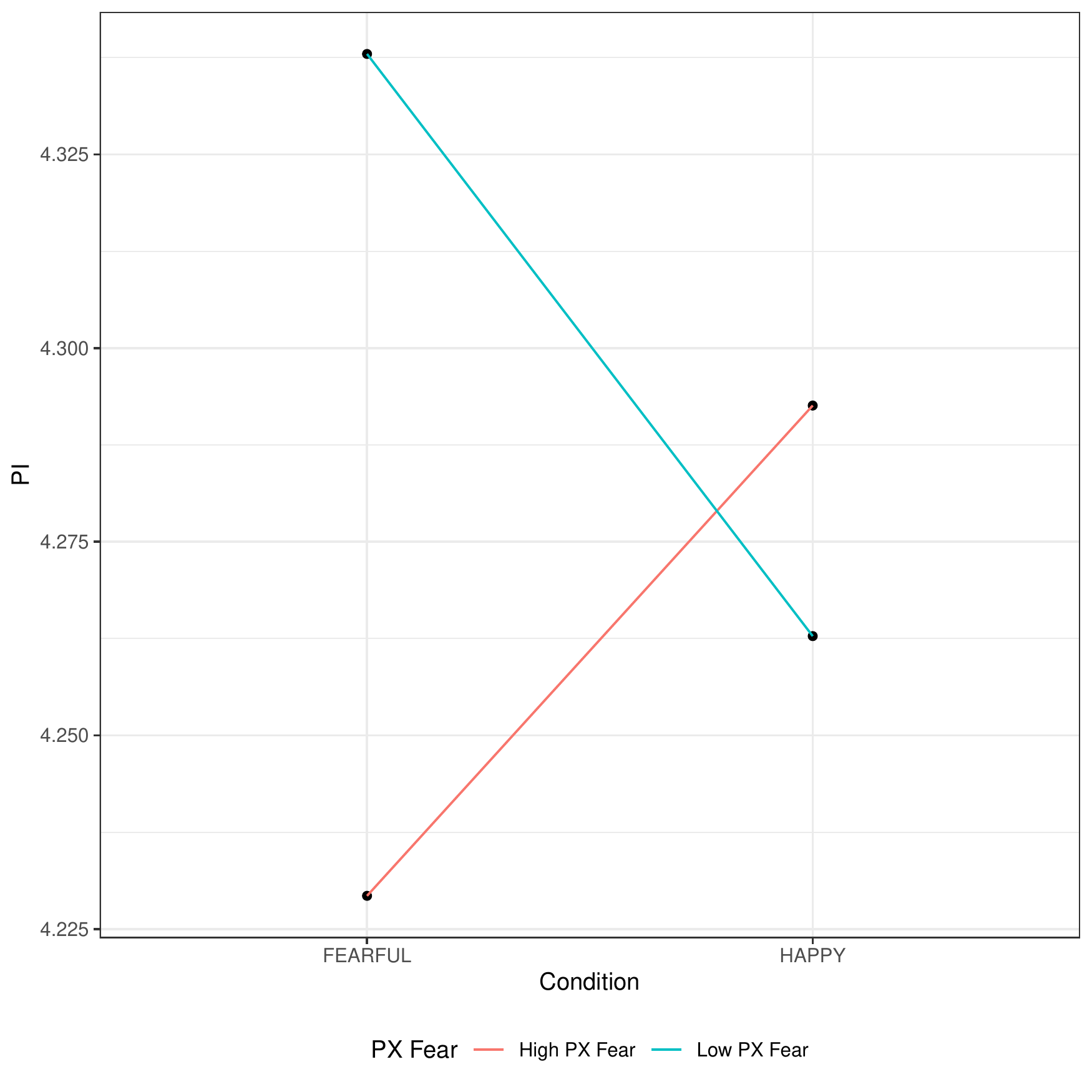} 
\label{fig:interactPIFear}
\end{minipage}
}~
\subfloat[TI/Fear]{
\begin{minipage}{0.3\textwidth}

\includegraphics[width=\linewidth]{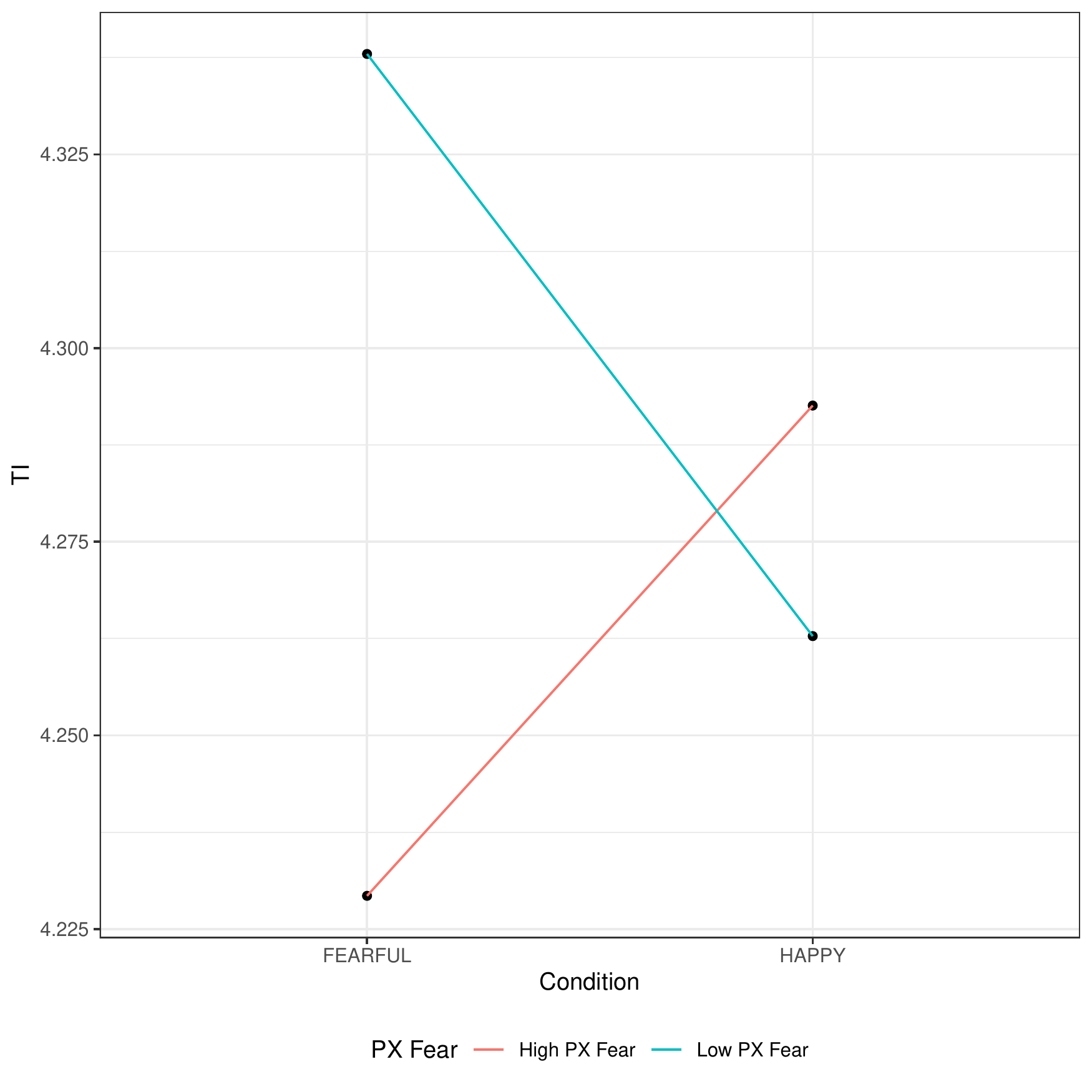} 
\label{fig:interactTIFear}
\end{minipage}
}

\subfloat[IDI/Joviality]{
\begin{minipage}{0.3\textwidth}

\includegraphics[width=\linewidth]{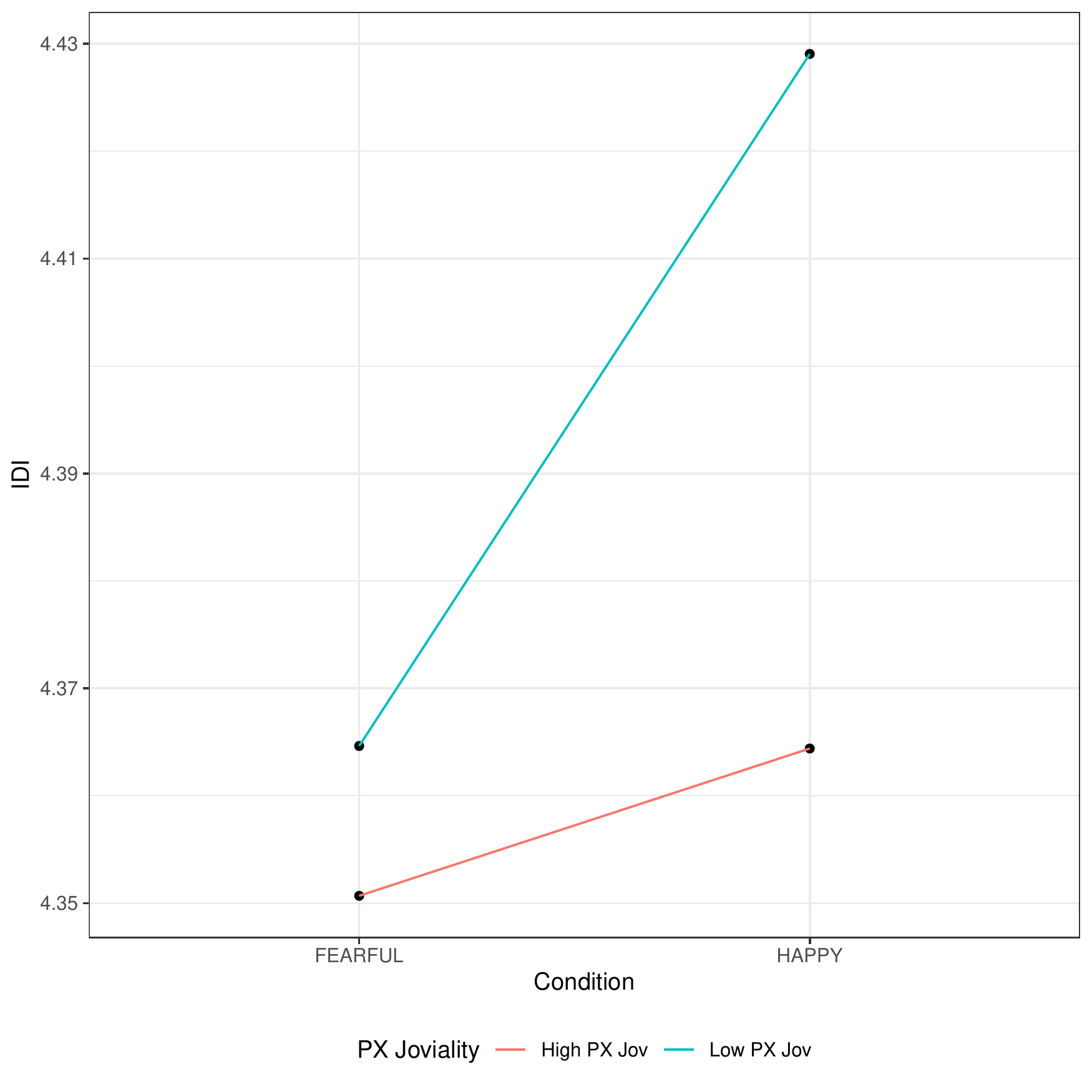} 
\label{fig:interactIDIJoviality}
\end{minipage}
}~
\subfloat[PI/Joviality]{
\begin{minipage}{0.3\textwidth}

\includegraphics[width=\linewidth]{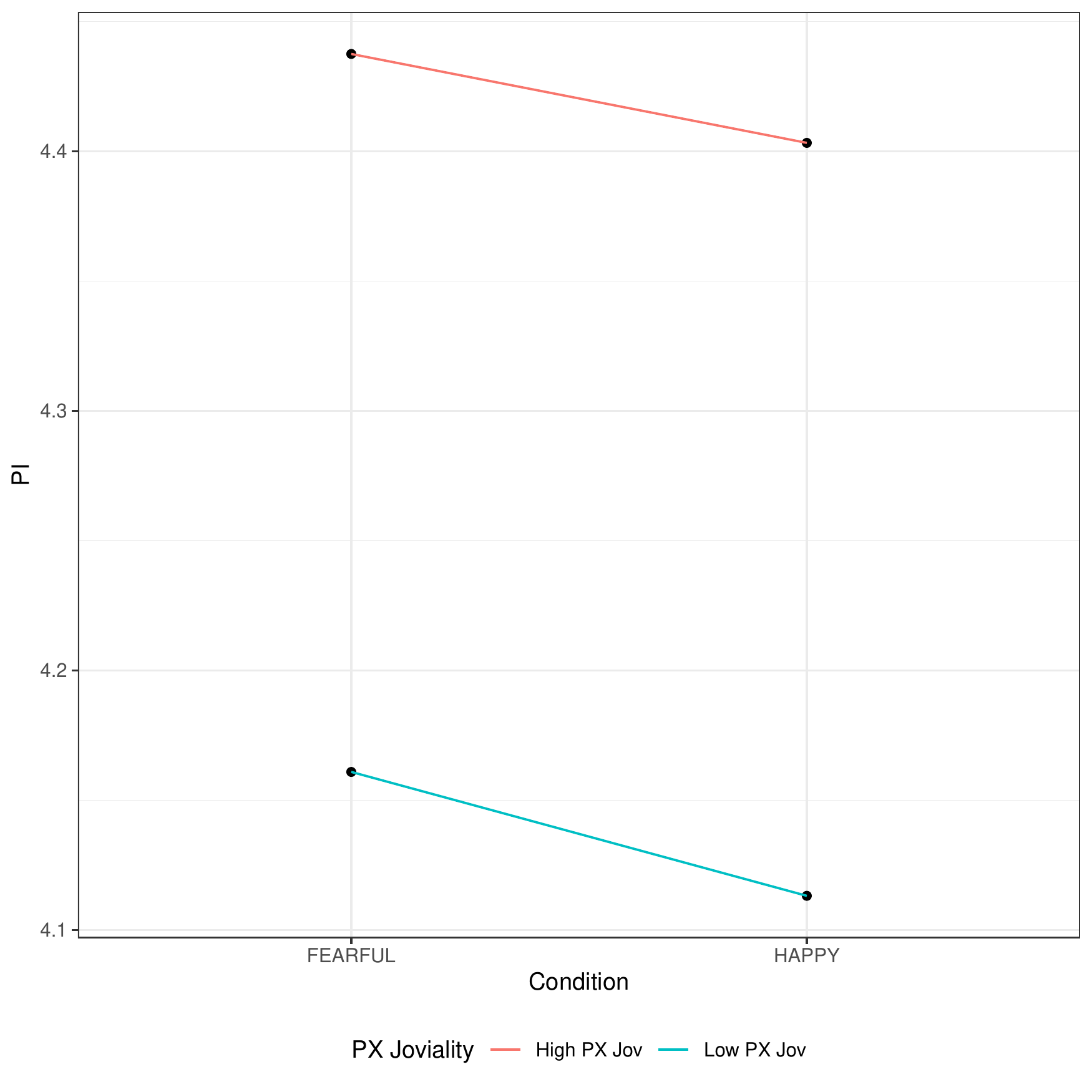} 
\label{fig:interactPIJoviality}
\end{minipage}
}~
\subfloat[TI/Joviality]{
\begin{minipage}{0.3\textwidth}

\includegraphics[width=\linewidth]{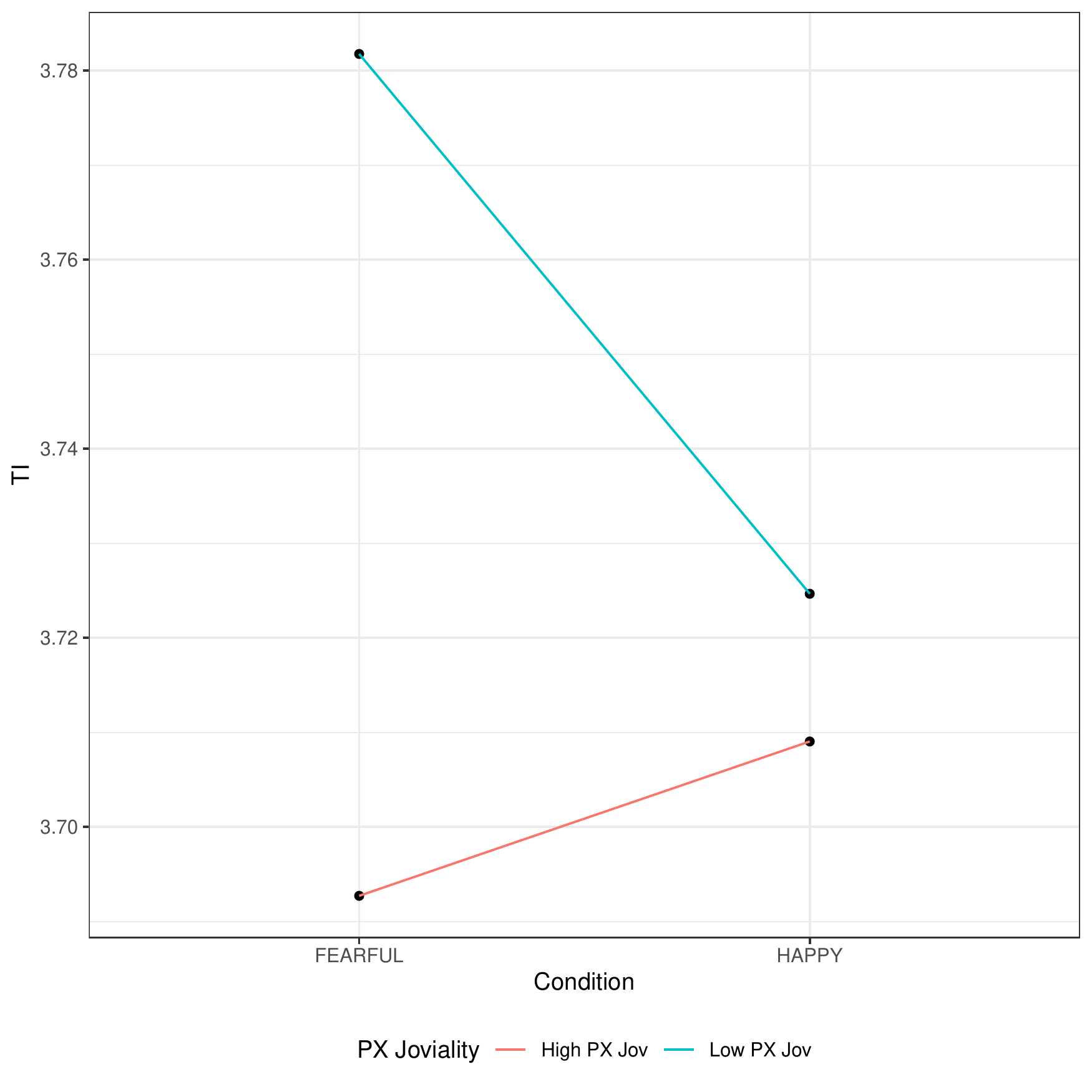} 
\label{fig:interactTIJoviality}
\end{minipage}
}
\caption{Interaction plots of PBI Subscales (IDI, PI, TI) on PX Affects and Condition.}
\label{fig:interactsOverall}
\end{figure*}
}

\newcommand{\forestPlotOverall}{
\begin{figure*}[htb]
\processifversion{DocumentVersionSTAST}{\vspace{-2.5cm}}

\includegraphics[width=\linewidth]{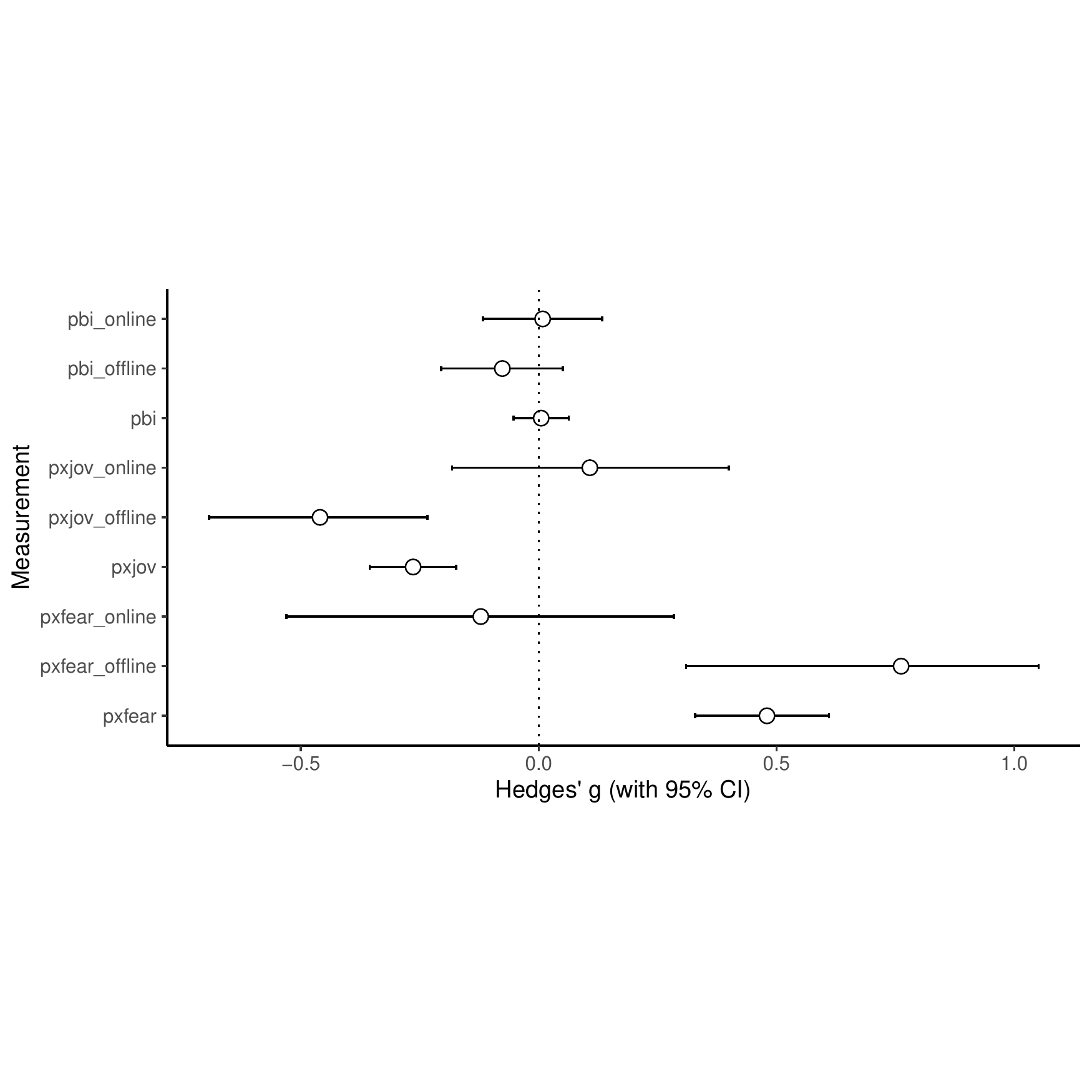} 
\processifversion{DocumentVersionSTAST}{\vspace{-2.5cm}}
\caption{Forest plot over manipulation check and PBI response variable, across datasets.}
\label{fig:forestPlotOverall}
\end{figure*}
}
\newcommand{\forestPlotPBIs}{
\begin{figure*}[htb]
\processifversion{DocumentVersionSTAST}{\vspace{-2.5cm}}

\includegraphics[width=\linewidth]{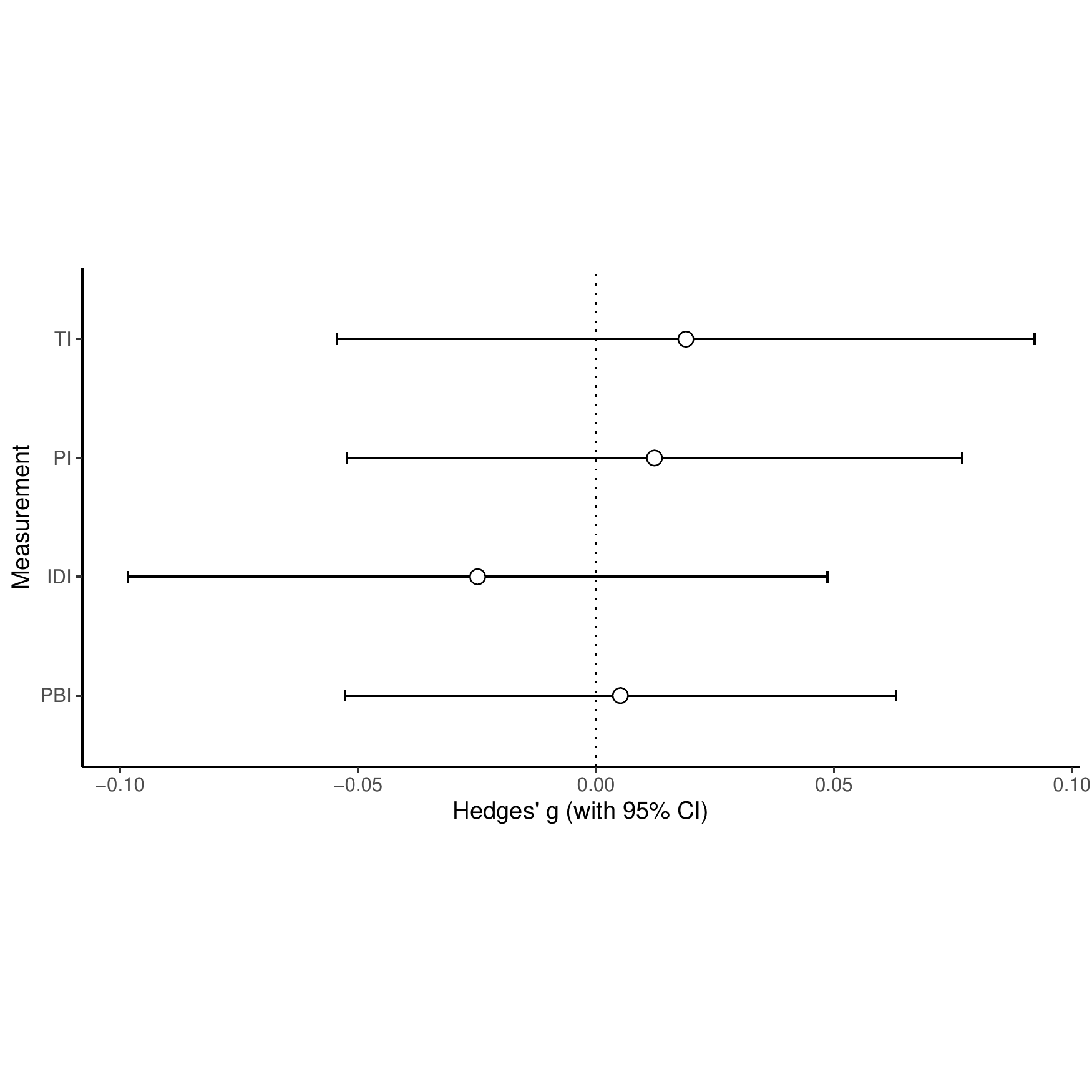} 
\processifversion{DocumentVersionSTAST}{\vspace{-2.5cm}}
\caption{Forest plot on PBI and its sub-scales for the combined dataset.}
\label{fig:forestPlotPBIs}
\end{figure*}
}
\newcommand{\forestPlotPBIsOverall}{
\begin{figure*}[htb]
\processifversion{DocumentVersionSTAST}{\vspace{-2.5cm}}

\includegraphics[width=\linewidth]{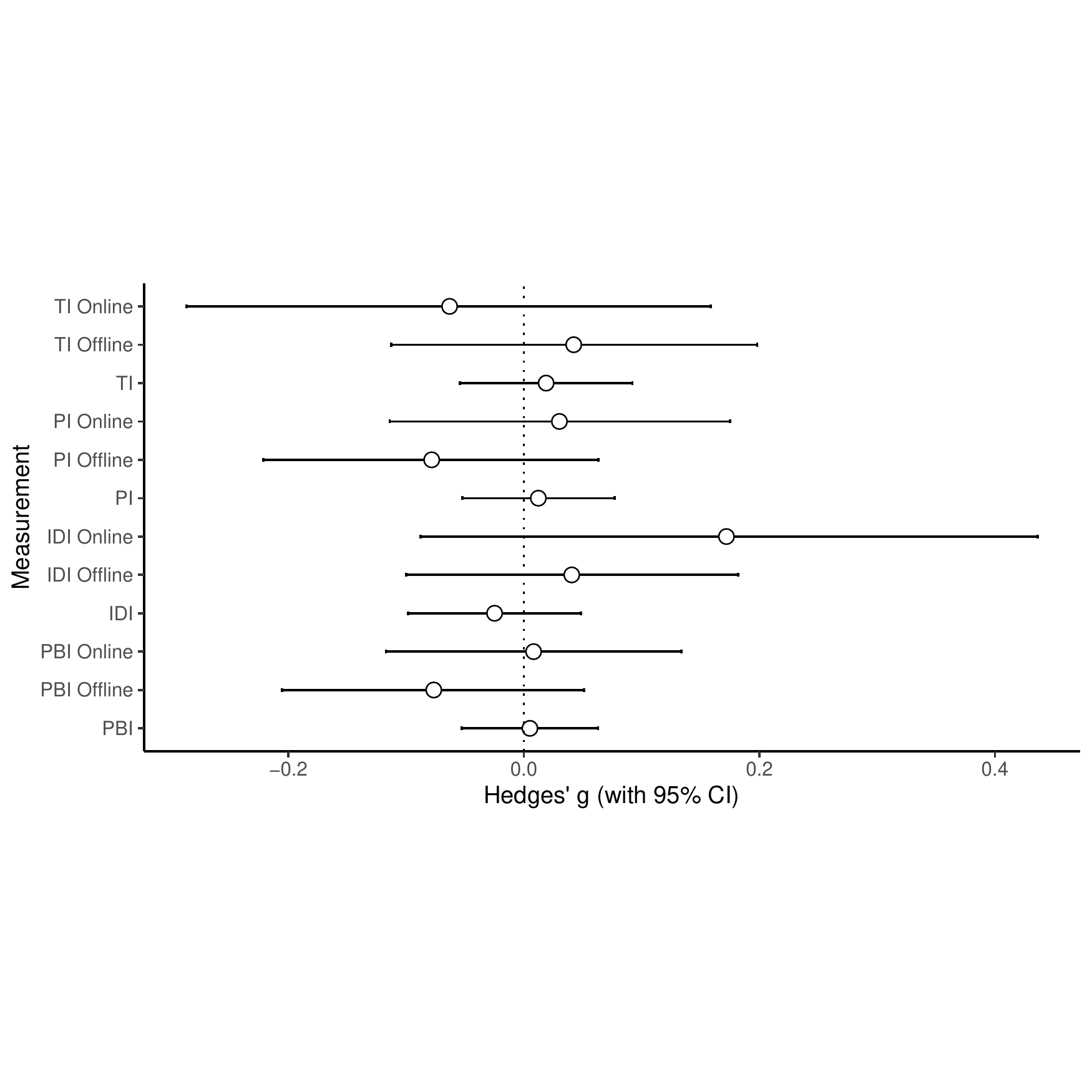} 
\processifversion{DocumentVersionSTAST}{\vspace{-2.5cm}}
\caption{Forest plot on PBI and its sub-scales for the combined dataset.}
\label{fig:forestPlotPBIsOverall}
\end{figure*}
}
\newcommand{\forestPlotPBIsNeutral}{
\begin{figure*}[htb]
\processifversion{DocumentVersionSTAST}{\vspace{-2.5cm}}

\includegraphics[width=\linewidth]{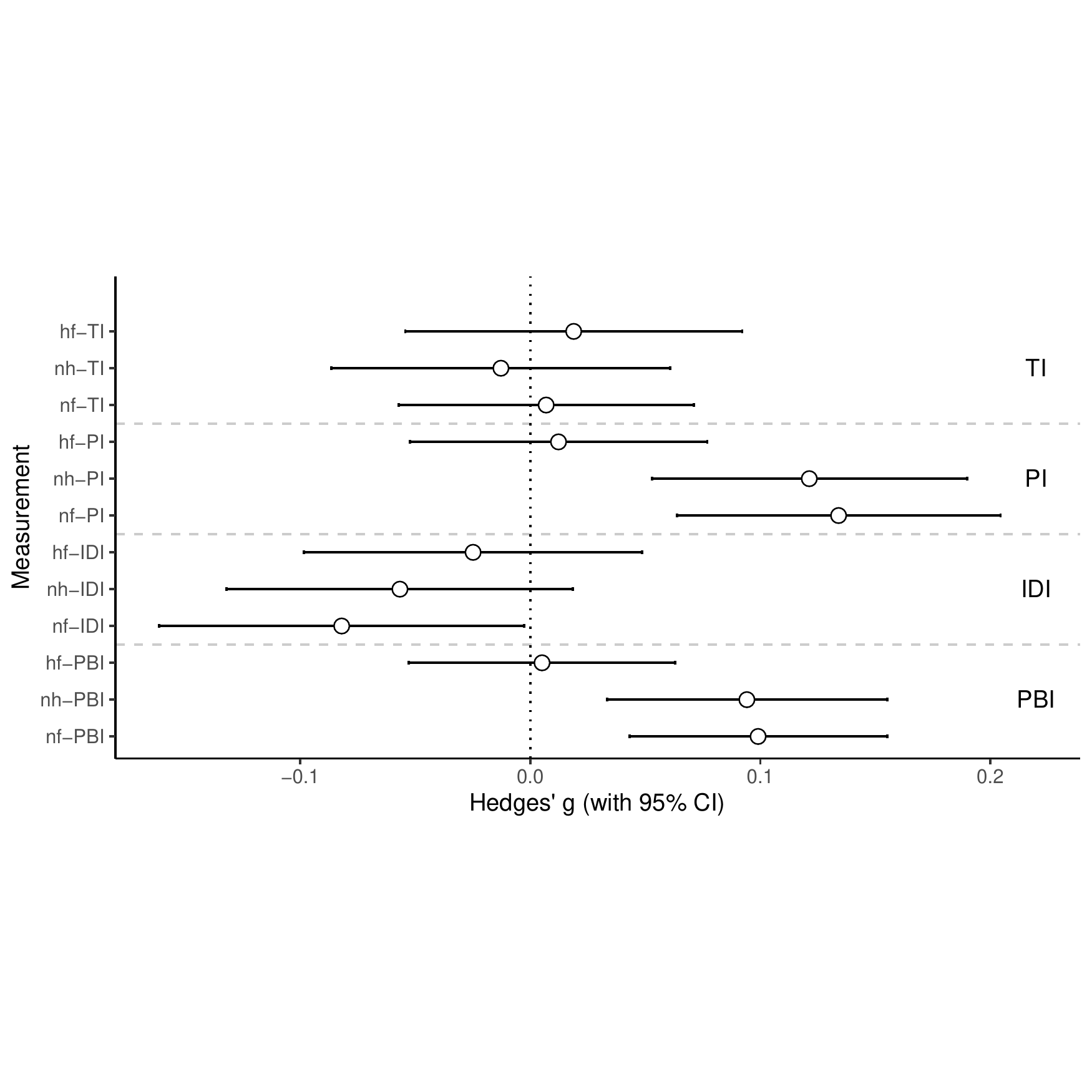} 
\processifversion{DocumentVersionSTAST}{\vspace{-3cm}}
\caption{Forest plot on PBI and its sub-scales for the combined dataset.}
\label{fig:forestPlotPBIsNeutral}
\end{figure*}
}


\date{}

\begin{DocumentVersionTR}
\title{Investigation of the Effect of Incidental Fear Privacy Behavioral Intention (Technical Report)\thanks{Open Science Framework Pre-Registration: \protect\url{https://osf.io/c3jy8/}. This is the technical report version of a paper to appear in proceedings of the 9\textsuperscript{th} International Workshop on Socio-Technical Aspects in Security (STAST'2019). LNCS 11739, Springer Verlag, 2019, pp. 181--204.}}
\end{DocumentVersionTR}

\begin{DocumentVersionTR}
\author{Uchechi Phyllis Nwadike\\Newcastle University, UK
\and
Thomas Gro{\ss}\\Newcastle University, UK}
\end{DocumentVersionTR}


\maketitle

\input{abstract.tex}

\input{intro.tex}

\input{background.tex}

\input{related.tex}

\input{aims.tex}

\input{method.tex}


\section{Results}

\begin{DocumentVersionTR}
\input{sample.tex}

We offer the descriptive statistics for the conditions \textsf{happy} and \textsf{fearful} across the three samples (Lab, MTurk, and Prolific) in Tables~\ref{tab:descriptivesOffline}, \ref{tab:descriptivesOnline}, and~\ref{tab:descriptivesProlific}, respectively.

\descriptivesOffline

\descriptivesOnline

\descriptivesProlific

\end{DocumentVersionTR}

\subsection{Manipulation Check: PANAS-X}

\paragraph{Assumptions.}
Both the differences between \textsf{happy} and \textsf{fearful} conditions on PANAS-X joviality and fear were not normally distributed,
$W = 0.94, p < .001$ and $W = 0.83, p < .001$ respectively.
We thereby choose to use a Wilcoxon Signed-Rank test.

\paragraph{Success of the Fear/Happiness Manipulations.}
The  fear reported by participants was statically significantly greather in the \textsf{fearful} condition ($M = 1.57$, $\vari{SD} = 0.77$) than in the \textsf{happy} condition ($M = 1.26$, $\vari{SD} = 0.49$), $V = 18286, p < .001$, 

The joviality of participants was statistically significantly less in the \textsf{fearful} condition ($M = 2.24$, $\vari{SD} = 1.1$) than in the \textsf{happy} condition ($M = 2.54$, $\vari{SD} = 1.17$), $V = 10398.5, p < .001$, Hedges' $g_{\const{av}} = -0.26$, 95\% CI $[-0.36, -0.17]$.

Hence, we reject the null hypotheses \const{H_{\vari{MC},\const{fear}, 0}} and \const{H_{\vari{MC},\const{jov}, 0}} and, thereby, consider the affect manipulation with the chosen video stimuli successful.

Figure~\ref{fig:plotTrigraphMC} contains a trigraph overview of all effects between all conditions, for joviality and fear respectfully.

\plotTrigraphMC

\subsection{Privacy Behavioral Intention}
\paragraph{Assumptions.}
The differences between \textsf{happy} and \textsf{fearful} conditions on Privacy Behavioral Intention (PBI) were not normally distributed,
$W = 0.94, p < .001$. We made the same observation for affect-neutral comparisons as well as sub-scales. We will use a Wilcoxon Signed-Rank test for the comparisons between conditions.

\paragraph{Differences Between Conditions.}
The PBI of the \textsf{fearful} condition ($M = 4.18$, $\vari{SD} = 0.9$) was significantly greater than in the \textsf{neutral} condition ($M = 4.1$, $\vari{SD} = 0.85$), $V = 28284.5, p_{\const{MC}(12)} < .001$, Hedges' $g_{\const{av}} = 0.1$, 95\% CI $[0.04, 0.16]$.

Similarly, the PBI of the \textsf{happy} condition ($M = 4.18$, $\vari{SD} = 0.9$) was significantly greater than in the \textsf{neutral} condition ($M = 4.1$, $\vari{SD} = 0.85$), $V = 29363.5, p_{\const{MC}(12)} < .001$, Hedges' $g_{\const{av}} = 0.09$, 95\% CI $[0.03, 0.16]$.

\emph{However}, PBI was \emph{not} statistically significantly different between \textsf{fearful} condition ($M = 4.18$, $\vari{SD} = 0.9$) and \textsf{happy} condition ($M = 4.18$, $\vari{SD} = 0.9$), $V = 23314.5, p_{\const{MC}(12)} = 1.000$, Hedges' $g_{\const{av}} = 0.01$, 95\% CI $[-0.05, 0.06]$.

Consequently, we reject the null hypotheses  \const{H_{\const{nf},\const{PBI}, 0}} and  \const{H_{\const{nh},\const{PBI}, 0}}.  However, we were not able to reject the null hypothesis  \const{H_{\const{hf},\const{PBI}, 0}}.

We report the difference between conditions first in the trigraph plot of Figure~\ref{fig:plotTrigraphPBI} on PBI and its subscales. Then we offer a traditional forest plot on the same data, with effect sizes grouped by DV scale (Figure~\ref{fig:forestPlotPBIsNeutral}).

\plotTrigraphPBI

\forestPlotPBIsNeutral

\subsection{PBI Sub-Scales}
We focus our attention on the PBI sub-scale comparisons that are likely statistically significant based on the effect sizes and confidence intervals reported in Figure~\ref{fig:forestPlotPBIsNeutral}.

The protection intention (PI) of the \textsf{fearful} condition ($M = 4.28$, $\vari{SD} = 1$) was significantly greater than of the \textsf{neutral} condition ($M = 4.15$, $\vari{SD} = 0.92$), $V = 26202.5, p_{\const{MC}(12)} < .001$, Hedges' $g_{\const{av}} = 0.13$, 95\% CI $[0.06, 0.2]$.

PI of the \textsf{happy} condition ($M = 4.27$, $\vari{SD} = 1$) was significantly greater than of the \textsf{neutral} condition ($M = 4.15$, $\vari{SD} = 0.92$), $V = 25749.5, p_{\const{MC}(12)} < .001$, Hedges' $g_{\const{av}} = 0.12$, 95\% CI $[0.05, 0.19]$.

Hence, we reject the null hypotheses \const{H_{\const{nf},\const{PI}, 0}} and  \const{H_{\const{nh},\const{PI}, 0}}, but failed to reject the null hypothesis \const{H_{\const{hf},\const{PI}, 0}}.

Finally, the significance of the IDI difference between \textsf{neutral} and \textsf{fearful} condition is in question (especially after MCC). And, indeed, we did not find a statistically significant difference between the IDI scores of these conditions, $V = 9504.5, p_{\const{MC}(12)} = .975$, $g_{\const{av}} = -0.08$, 95\% CI $[-0.16, 0]$.

For null hypothesis \const{H_{\const{nf},\const{IDI}, 0}} as well as the remaining sub-scale null hypotheses, we consider them as not rejected.

\subsection{Interactions}
It deserves closer attention why the main condition (happiness-fear) is shows a lower effect size than comparisons between the neutral and affect-induced states. To shed light on this situation we compute interaction plots on dichotomized PX fear and happiness scores (high/low). 

\paragraph{PBI Interactions.} Figure~\ref{fig:interactsPBI} considers the interactions for the main DV \textsf{pbi}.
Sub-Figure~\ref{fig:interactPBIFear} clearly shows the cross-over interaction between happiness and fear condition on the experienced fear in high or low levels of PX fear. Joviality (\ref{fig:interactPBIJoviality}) does not show any interaction.

Whereas participants with high fear in the fear condition exhibit a lower PBI score, participants with high fear in the happiness condition show a higher PBI score. Vice versa, participants with low fear show a higher PBI score in the fear condition and show a lower PBI score in the happiness condition.

\interactsPBI

\paragraph{Sub-Scale Interactions.} 
We face a complex situation in the interactions on sub-scales, displayed in Figure~\ref{fig:interactsOverall}.
There are varying degrees of interactions. 

Information Disclosure Intention (IDI) is impacted by interactions in opposing directions, yet not crossing over (\ref{fig:interactIDIFear} and \ref{fig:interactIDIJoviality}).

Fear yields a cross-over interaction on Protection Intention (PI) (\ref{fig:interactPIFear}).
The impact of joviality on PI yield no interaction (\ref{fig:interactPIJoviality}).
This sub-scale shows the clearest difference between impact of fear and joviality on sub-scale.

Fear has a cross-over interaction on Transaction Intention (TI) (\ref{fig:interactTIFear}).
Joviality shows a milder interaction on TI in the same direction (\ref{fig:interactTIJoviality}).

\interactsOverall

\begin{DocumentVersionTR}
\forestPlotPBIs

\forestPlotPBIsOverall
\end{DocumentVersionTR}

 
\input{discussion.tex}
 
\input{limitations.tex}

\input{conclusion.tex}

\input{acknowledgment.tex}

\balance

\bibliographystyle{IEEEtran}
\bibliography{references}

\end{document}

%% file: abstract.tex
\renewcommand{\abstractname}{Structured Abstract}
\begin{abstract}
\noindent{\bfseries Background.}
Incidental emotions users feel during their online activities may alter their privacy behavioral intentions.

\noindent{\bfseries Aim.} 
We investigate the effect of incidental affect (fear and happiness) on privacy behavioral intention.

\noindent{\bfseries Method.} 
We recruited $330$ participants for a within-subjects experiment in three random-controlled user studies. The participants were exposed to three conditions \textsf{neutral}, \textsf{fear}, \textsf{happiness} with standardised stimuli videos for incidental affect induction. Fear and happiness were assigned in random order. The participants' privacy behavioural intentions (PBI) were measured followed by a Positive and Negative Affect Schedule (PANAS-X) manipulation check on self-reported affect.  
The PBI and PANAS-X were compared across treatment conditions.

\noindent{\bfseries Results.} 
We observed a statistically significant difference in PBI and Protection Intention in neutral-fear and neutral-happy comparisons. However across fear and happy conditions, we did not observe any statistically significant change in PBI scores.

\noindent{\bfseries Conclusions.} 
We offer the first systematic analysis of the impact of incidental affects on Privacy Behavioral Intention (PBI) and its sub-constructs. We are the first to offer a fine-grained analysis of neutral-affect comparisons and interactions offering insights in hitherto unexplained phenomena reported in the field.
\end{abstract}

%% file: intro.tex
\section{Introduction}

Online privacy behaviors, though not limited to requesting for personal contact information to be removed from mailing lists, keeping passwords safe, use of strict privacy settings~\cite{mcguinness2018information} such as can be observed as users take deliberate steps to protect their personal details while browsing on the Internet. Despite the best intentions to avoid sharing personal information, users still get influenced by different factors to disclose same. These factors include incentives, loyalty points, privacy concerns to mention a few. There is sparse information on the effect of emotion on privacy behavior despite existing literature in psychology and economics highlighting the role of emotion in human behavior~\cite{loewenstein2000emotions} and decision making~\cite{loewenstein2000emotions}.
Though preliminary studies~\cite{coopamootoo2017wip,wakefield2013influence} have been conducted on the effect of fear, happy, anger affect states on privacy behavioral intentions, however no comprehensive study has been conducted to investigate the effect of the relationship between neutral, fear and happy affect states on privacy behavioral intentions. 
Researchers such as Wakefield~\cite{wakefield2013influence}, Li~\cite{li2011role} or Coopamootoo and Gro{\ss}~\cite{coopamootoo2017privacy} have called for further investigation to extend the existing knowledge on the effect of fear and happy affect states on privacy behavioural intentions. 

In this paper we contribute a comprehensive study which systematically compares PANAS-X and PBI scores across neutral, fear and happy conditions and explores the relationship between affect states and privacy behavioral intentions.

Given that measurement of actual privacy behavior is difficult to achieve within a laboratory setting, privacy behavioral intentions are frequently used as a proximal measure~\cite{gerber2018explaining,li2017resolving}. Hence in our user study we will measure privacy behavioral intentions instead of privacy behavior.                                                                                                                                                                                                                                                                                                                                                                                                                                                                                                                                                                                                                                                                                         
The aim of the user studies discussed in this paper is to investigate the effect of incidental affect states of induced fear and happy affect states on privacy behavioral intention. This will contribute towards bridging the highlighted research gap, extend the existing research knowledge and provide empirical evidence that would be useful for further research.

%% file: background.tex
\section{Background}

\subsection{Affect, Emotion, and Mood}

The terms mood, emotion and affect have been used interchangeably in literature in the past, however referred to a range of emotional intensity and also reflect fundamental differences including duration, frequency, intensity and activation pattern. 
The terms \emph{affect} or \emph{affective states} are often used to describe the experience of emotion or feeling, in general, going back to an early definition attempt of Scherer~\cite{scherer1984nature}.
The terms \emph{core affect}, \emph{emotion} and \emph{mood} have been differentiated further in subsequent years. Ekkekakis~\cite{ekkekakis2013measurement} made one of the most recent attempts to summarize the emergent consensus of the field as well as the differences between these constructs (cf. \cite[Ch. 7]{ekkekakis2013measurement} for a detailed analysis).

We note that the impact of affect on another task led to a classification of \emph{incidental affect}, that is, affect independent from the task at hand, and \emph{integral affect}, that is, affect related to the task at hand.  The differences between incidental and integral affect have received attention in psychology research~\cite{vastfjall2016arithmetic,lerner2000beyond,han2007feelings}.

Let us consider these terms and converge on a definition for this paper.
\subsubsection{Core Affect.}
\begin{figure*}[tb]
\centering
\includegraphics[keepaspectratio,width=0.5\linewidth]{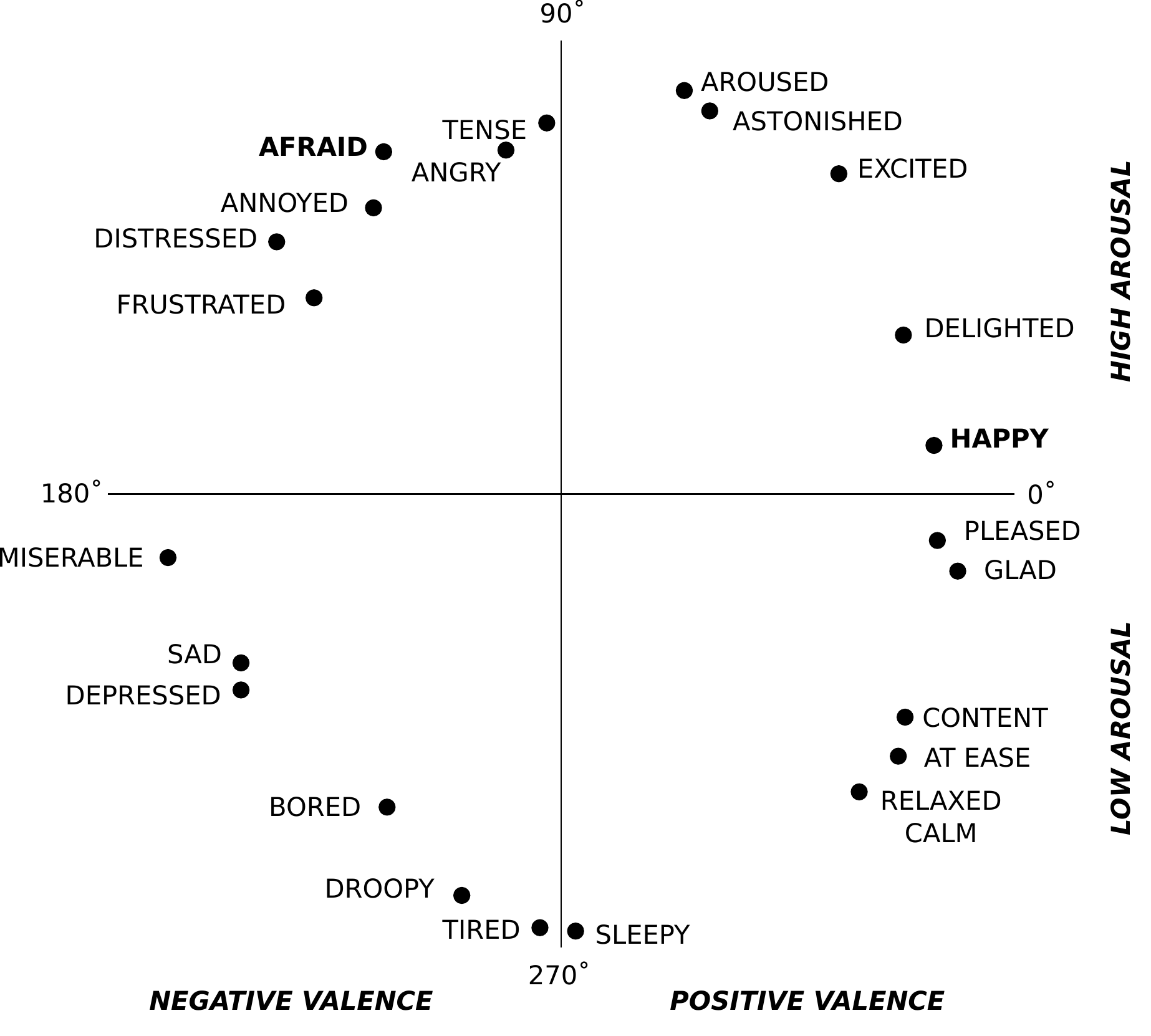}
\caption{Circumplex model of affect, adapted from Russel~\cite{russell1980circumplex}. Note that fear is classified as negative-valence/high-arousal ($110^{\circ}$, labeled by Russel as ``afraid'') and happiness is classified as positive-valence/high-arousal ($10^{\circ}$, labeled as ``happy''). PANAS-X includes the high-arousal items delighted ($30^{\circ}$) and excited ($50^{\circ}$) in its joviality scale.}
\label{fig:circumplex}
\end{figure*}

Following the discussion by Ekkekakis~\cite{ekkekakis2013measurement}, we perceive core affect is a broader concept than mood or emotion.
 Russell~\cite{russell2003core} defined (core) affect as the specific quality of goodness or badness experienced as a feeling state (with or without consciousness). We include his circumplex model of affect in Figure~\ref{fig:circumplex}. Affect states can be triggered spontaneously by memories, exposure to stimuli, perception of one's immediate environment~\cite{coan2007handbook,vastfjall2016arithmetic,slovic2005affect}.  
Subsequently, Russel and Feldman Barrett~\cite{russell1999core} offered an updated definition ``core affect is defined as a neuro-physiological state consciously accessible as a simple primitive non-reflective feeling most evident in mood and emotion but always available to consciousness'' also consulted by Ekkekakis~\cite{ekkekakis2013measurement}.
The feeling being non-reflective has been pointed out as a critical attribute.

\subsubsection{Emotion.}
Lazarus~\cite{lazarus1991emotion} defined \emph{emotion} ``as a complex reaction of a person arising from appraisals and outcomes of self-relevant interactions with the environment, which could result in states of excitement, direction of attention, facial expressions, action tendencies, and behavior.''
Ekkekakis~\cite[Ch. 7]{ekkekakis2013measurement} points out elaborating on the discussion by Lazarus that: ``Because emotions are elicited \emph{by} something, are reactions \emph{to} something, and are generally \emph{about} something, the cognitive appraisal involved in the person-object transaction is considered a defining element.'' (emphasis by the Ekkekakis).

\subsubsection{Mood.}
Mood is typically differentiated from emotion by intensity and duration. Lazarus~\cite{lazarus1991emotion} discussed \emph{mood} as follows: ``While moods are usually less anchored to specific stimulus conditions and perhaps of longer duration and lesser intensity than emotions, may also be distinguished in terms of specific content.''  
Similarly, Ekkekakis~\cite[Ch. 7]{ekkekakis2013measurement} discusses that moods are ``diffuse and global.''

\subsection{Affect Elicitation}
This refers to the process of engaging study participants in specific tasks with the sole purpose of drawing out the required affect state of interest from individuals. Given the brief span associated with affect states it is difficult to determine how long a treatment should be in order to obtain and ascertain the affect's intensity and  sustained effect  throughout the user study duration. Emotion researchers have recommended that the inducement process should not last longer than minutes~\cite{qiao2017transient}.

The methods used to elicit or induce affect states include the use of standardized stimulus in form of audio, video, autobiography recall, vignettes and pictures~\cite{siedlecka2018experimental}. These methods are not used isolation rather the authors recommend to use a combination of techniques. They proposed that different affect inducing methods have different success rates with evoking different affect states within the laboratory's setting~\cite{siedlecka2018experimental}.  The methods are further discussed below. 

\subsubsection{Visual Stimuli.}
This refers to the use of visually stimulating images such as video clips as stimuli. 
These materials can contain either evocative or non evocative elements which can induce specific affect states.
 Types of visual stimuli include pictures (e.g., gruesome images, a sunset over a calm sea) or  films (e.g., defined scenes from horror or comedy films).
 
To induce fear, psychologists~\cite{coan2007handbook} have established standardized scenes from ``\emph{The Shining}'' or ``\emph{The Silence of the Lambs}.'' To induce happiness, similarly, the restaurant scene from ``\emph{When Harry meets Sally}'' was evaluated. As a standard procedure, these stimuli videos are precisely defined and systematically evaluated for their impact on the participants' affective state.
In the studies discussed later in this chapter, we selected the use of standardized stimulus video films. This technique was considered the most effective in inducing affect states with large effect sizes.

\subsubsection{Autobiographical Recall.}
Participants are requested to recollect (and at times write) about real life evoking events from their past when they experienced a particular affect state the researchers are interested in. The researchers expect participants to relive the affect state felt at the time of the event or incident. Studies which involved the use of this method have recorded noticeable change in emotional effect in self reported responses and increased physiological responses such as heart rate, skin conductance~\cite{marci2007autonomic}. 

\subsection{Affect Measurement.}

The methods and tools used in measuring affect states can be classified into two categories, namely psychometric self-report instruments and psycho-physiological measurements. 

Self report tools involve the use of scales, such as PANAS-X, which require input from the subject, reporting how he or she feels with respect to a defined timeframe (e.g., in this moment).
In this study, we have selected the Positive and Negative Affect Schedule (PANAS-X)~\cite{watson1999panas} as manipulation check of choice. While largely considered an effective measurement instrument~\cite{watson1988development,crawford2004positive}, Ekkekakis~\cite[Ch. 12]{ekkekakis2013measurement} pointed out that the PANAS-X items include core affects, emotions, as well as moods in the terminology set out above. We also note that the sub-scales on \emph{fear} and \emph{joviality} we use in this study would be considered an emotion in this terminology, infused with high negative and positive core affect, respectively. We note that in recent years, there was further criticism of the factorial structure and theoretical underpinnings of PANAS-X, especially when it comes to measuring low-activation states (outside of the scope of this study). 

Psycho-physiological measurement tools do not require any subjective input rather it involves the measurement of physiological responses after exposure to a given stimulus. These responses could be facial expressions, heart beat, skin conductance, pupil movement to mention a few. While there has been supportive evidence for those tools, there is also criticism in constructivist views on emotions that physiological states are not necessarily indicative of specific emotions.

%% file: related.tex
\section{Related Work}
To the best of our knowledge, we were the first to present an affect induction experiment designed to explore the effect of affect states on privacy behavioral intention (PBI)~\cite{nwadike2016evaluating}. Our paper reports pilot studies and describes a design template to investigate the impact of happy and fear affect states on PBI. 
Subsequently, Coopamootoo~\cite{coopamootoo2017wip} and Fordyce et al.~\cite{fordyce2018investigation} adopted said template in their research. So do we. Fordyce et al. applied the method to password choice.

The closest work to this study is Coopamootoo's work-in-progress (WIP) paper~\cite{coopamootoo2017wip}. Both experiments are measuring incidental affect as proposed by Nwadike et al.~\cite{nwadike2016evaluating} initially.
for which we notice a number of differences:
\begin{inparaenum}[(i)]
\item While this study is only concerned with the impact of affect on PBI, the study by Coopamootoo also considered general self-efficacy as a human trait.
\item While we induced incidental affects with standardized stimuli films, Coopamootoo used autobiographical recall of
	emotive events. 
\item While Coopamootoo uses tone analysis of participant-written text as predictor, we use PANAS-X as psychometric instrument. Our impression is that the tone analysis only measures the tone of a text input, but does not constitute a psychometric measurement of the current affective state of a participant.
\end{inparaenum}

We noticed that the correlations reported in the WIP paper~\cite[\S 5.3.4]{coopamootoo2017wip} are trivial to small ($r \leq 0.3$), with Protection Intention (PI) having the greatest reported correlation with self-efficacy ($r = 0.3$). 

Coopamootoo's causal hypothesis $\const{H}_{\const{C}, 1}$ was declared as ``emotion [fear/happiness] and self-efficacy impact privacy intentions.'' The relations reported are
\begin{inparaenum}[(i)]
\item emotions impact trait self-efficacy (negatively) and 
\label{clause:aff-SE}
\item self-efficacy impacts PI (positively).
\end{inparaenum}
Strikingly, for relation~(\ref{clause:aff-SE}) the corresponding study~\cite[Study 2, \S 4.2]{coopamootoo2017wip} administered the trait self-efficacy questionnaire in Step (b) \emph{before} the affect induction protocol (Step (c)). Hence, the induced emotion could \emph{not} have \emph{caused} self-efficacy changes.
Thus, we need to assume that Coopamootoo tried models with a number of sub-scales of PBI (4), different emotion scales (4) and with different SEM path models while not correcting for these multiple analyses. We consider the reported causal relation a false positive. 

%% file: aims.tex
\section{Aims}

\paragraph{Impact of Affect.} 
The study seeks to make a comparison of the influence of affect states on Privacy Behavioral Intention (PBI)~\cite{yang2009influence}.

\begin{researchquestion}[Impact of affect on PBI]
To what extent does Privacy Behavioral Intention (PBI)~\cite{yang2009influence} in the form of Information Disclosure Intention (IDI), Transaction Intention (TI) and Protection Intention (PI) change depending on induced incidental happy and fear states?
\end{researchquestion}

This research question decomposes into multiple statistical hypotheses iterated over dependent variables (\const{idi}, \const{ti}, \const{pi}) pair-wise compared across conditions (\const{neutral}, \const{happy}, \const{fear}). Hence, we obtain nine null and alternative hypotheses pairs for comparisons on: \const{neutral}--\const{happy} (\const{nh}), \const{neutral}--\const{fear} (\const{nf}), \const{happiness}--\const{fear} (\const{hf}). In addition to that, we investigate the pair-wise comparison of the combined \const{pbi} scores across conditions.

As primary analysis, we are most interested in, we consider the combined Privacy Behavioral Intention (PBI) \const{pbi} in the comparison between the \const{happy}--\const{fear} conditions.

 \const{H_{\const{hf},\const{pbi}, 0}} There is no difference in privacy behavioral intention (\const{pbi}) between cases with induced happiness and induced fear.
  \const{H_{\const{hf},\const{pbi}, 1}} Privacy behavioural intention \const{pbi} differs between the conditions \const{happy} and \const{fear} (\const{hf}).

The hypotheses are obtained iterating over
\begin{compactenum}
  \item Conditions Comparisons \vari{CC} :=
  \begin{compactenum}
     \item \const{neutral}--\const{happy} (\const{nh}), 
     \item \const{neutral}--\const{fear} (\const{nf}), 
     \item \const{happy}--\const{fear} (\const{hf})
   \end{compactenum}
   \item Dependent Variables \vari{DV} :=
   \begin{compactenum}
     \item \const{pbi}: Combined Privacy Behavioral Intention (PBI) score,
     \item \const{idi}: PBI Information Disclosure Intention (IDI) sub-scale score,
     \item \const{ti}: PBI Transaction Intention (TI) sub-scale score,
     \item \const{pi}: PBI Protection Intention (PI) sub-scale score.
   \end{compactenum}
\end{compactenum}
 \const{H_{\vari{CC},\vari{DV}, 0}} There is no difference in privacy behavioral intentions scores of scale \vari{DV} between conditions specified in comparison \vari{CC}.
  \const{H_{\vari{CC},\vari{DV}, 1}} Privacy behavioural intention scores of scale \vari{DV} differ between the conditions specified in comparison \vari{CC}.
Note that \vari{CC} and \vari{DV} are variables that take values $(\const{nh}, \const{nf}, \const{hf})$ and $(\const{pbi}, \const{idi}, \const{ti}, \const{pi})$ as specified above. They thereby define a test family of 12 alternative and null hypothesis pairs.

\paragraph{Manipulation Check.}
We use the Positive and Negative Affect Schedule (PANAS-X)~\cite{watson1999panas} as joint manipulation check \vari{MC} across the three studies..

A manipulation is considered successful if the following null hypotheses can be rejected.
 \const{H_{\vari{MC},\const{jov}, 0}} There is no difference in \vari{MC}-measured joviality between happy and fear conditions.
  \const{H_{\vari{MC},\const{jov}, 1}} The \vari{MC}-measured joviality differs between happy and fear conditions.

 \const{H_{\vari{MC},\const{fear}, 0}} There is no difference in \vari{MC}-measured fear between happy and fear conditions.
  \const{H_{\vari{MC},\const{fear}, 1}} The \vari{MC}-measured fear differs between happy and fear conditions.
The pre-registration defined the PANAS-X (\const{px}) measurement as authoritative.

\paragraph{Regression Model.}
We are interested in the relation between Privacy Behavioral Intention $(\const{pbi}, \const{idi}, \const{ti}, \const{pi})$ and measured affect PANAS-X (\const{pxjov} and \const{pxfear}).

\begin{researchquestion}[Relation of measured affect and PBI]
To what extent is there a linear relationship between the reported affect state (PANAS-X) and the PBI scales.
\end{researchquestion}

We consider the following hypotheses for an overall model, with canonical hypotheses for the respective predictors.
 \const{H_{\const{px},\const{pbi}, 0}} There is no linear relationship between measured affect and PBI scores.
  \const{H_{\const{px},\const{pbi}, 1}} There is a systematic linear relationship between measured affect and PBI scores.
We note that the overall PBI score relation is designated as primary hypothesis and the PBI-sub-scale relations designated as secondary hypotheses.

\begin{table*}[tb]
\centering
\footnotesize
\caption{Operationalization: Effect of Fear on Privacy Behavioral Intention.}
\label{tab:ops_fear}
\begin{tabular}{llll}
\toprule
                & Levels	    & Instrument & Intervention/Variable\\
\midrule
IV: Affect	& Fear 	    & Stimulus Video~\cite{ray2007emotion} & \emph{The Shining} \\
		& Happiness & 								&\emph{When Harry Met Sally} \\
\midrule
IV Check & Fear		     & PANAS-X~\cite{watson1999panas}    & \textsf{fear} \\
	        & Happiness 	      & 						    & \textsf{joviality} \\
\midrule
DV: Privacy Behavioral Intention &	   & PBI~\cite{yang2009influence} & \textsf{pbi} \\
\qquad \emph{Sub-Scales:} \\
\multicolumn{2}{l}{\qquad  Information Disclodure Intention}   &    & \textsf{idi}\\
\multicolumn{2}{l}{\qquad   Transaction Intention}   &    & \textsf{ti}\\
\multicolumn{2}{l}{\qquad   Protection Intention}   &    & \textsf{pi}\\
\bottomrule
\end{tabular}
\end{table*}

%% file: method.tex
\section{Method}
The studies had been pre-registered at Open Science Framework\footnote{\url{https://osf.io/c3jy8/}}. The tables and graphs were produced directly from the data with the \textsf{R} package \textsf{KnitR}. 

The statistical inferences conducted are two-tailed and operate on a significance level of $\alpha = .05$. We consider the per-condition pair-wise Wilcoxon Signed-Rank tests for Privacy Behavior Intention (PBI) and its sub-scales IDI, PI, and TI as a test family with $12$ elements. A Bonferroni-Holm multiple-comparison correction (MCC) is applied directly to the $p$-values reported, indicated as $p_{\const{MC}(12)}$. 

Even though the differences between DVs are at times not normally distributed, we use Hedges $g_{\const{av}}$ as unbiased dependent-samples effect size and its corresponding confidence interval for estimation purposes.

The three studies were conducted within online and offline settings: one laboratory and two online studies. We chose to run these studies in a combination of offline and online settings for two main reasons: because we wanted to measure facial expressions, which at the time could be done only within a lab setting; and also have access to a larger sample size.

\subsection{Experiment Design Evaluation}
\begin{figure*}[p]
\centering
\includegraphics[keepaspectratio,width=\textwidth, trim=1cm 0cm 1.3cm 0cm]{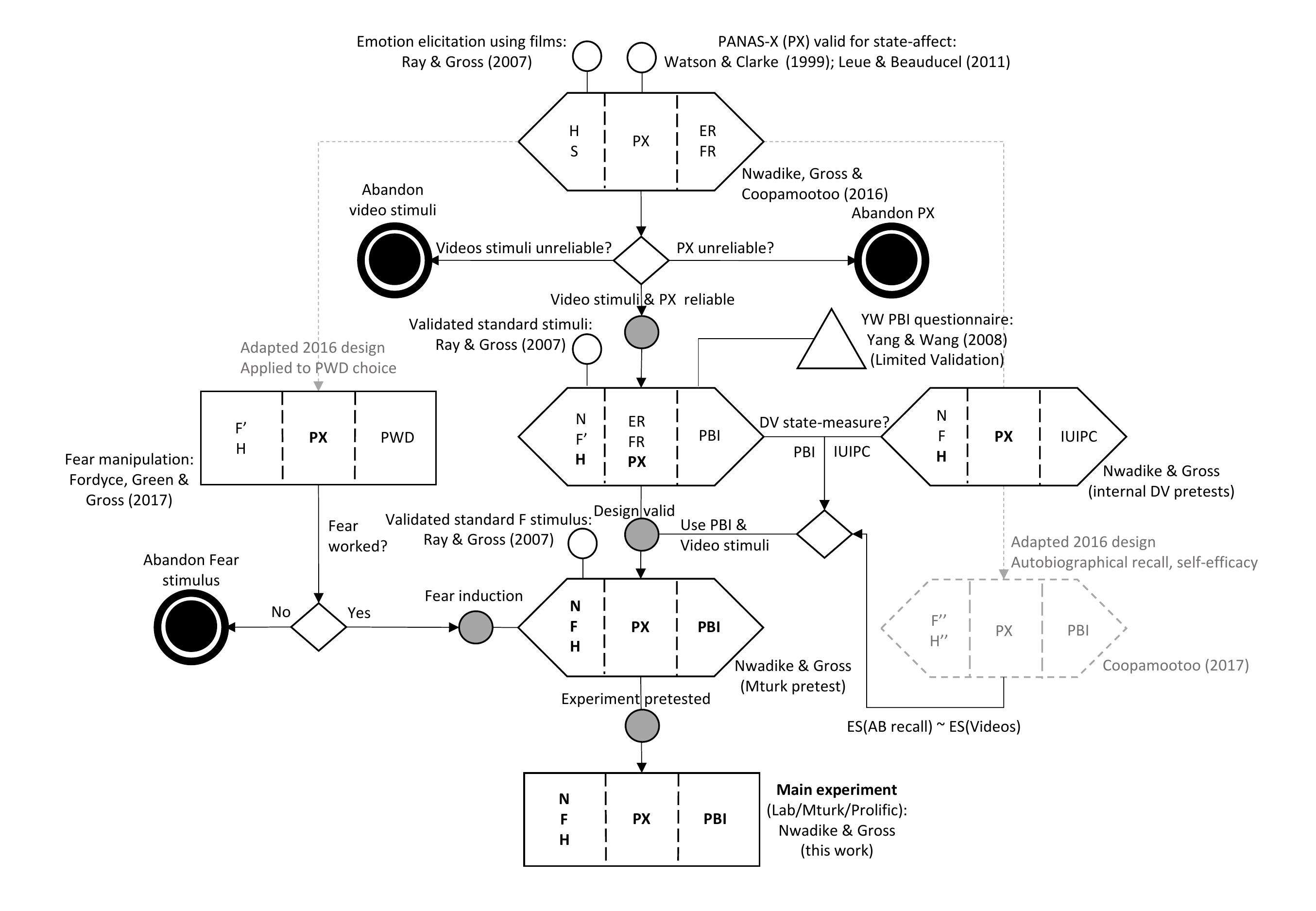}
\includegraphics[keepaspectratio,width=\textwidth]{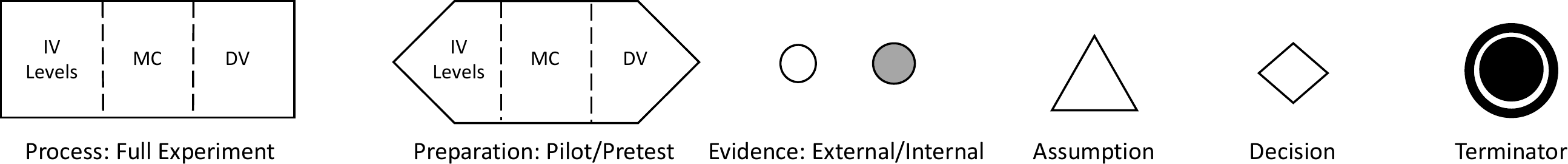}
\caption{Flowchart of the research process from 2016--2019, highlighting external evidence drawn upon, assumptions and design decisions made, as well as different pilots/pretests and external experiments informing this main study. 
We model Processes and Preparations as tripartite, consisting of 
1. Independent Variable (IV) Levels,
2. Manipulation Check, and
3. Dependent Variable (DV).
\textbf{IV Levels:} \textsf{N} = neutral (video ``\textit{Alaska's Wild Denali}'')~\cite{ray2007emotion}; \textsf{H} = happiness (video ``\textit{When Harry met Sally}'')~\cite{ray2007emotion}; \textsf{F} = fear (video 1 ``The Shining'')~\cite{ray2007emotion}; \textsf{F'} = fear (video 2 ``The Silence of the Lambs'')~\cite{ray2007emotion}; \textsf{H''} = happiness (autobiographic recall); \textsf{F''} = fear (autobiographic recall); \textsf{S} = sadness (video ``\textit{The Champ}'')~\cite{ray2007emotion}; \textbf{Manipulation Checks:} \textsf{PX} = PANAS-X~\cite{watson1999panas}; \textsf{ER} = Microsoft Emotional Recognition; \textsf{FR} = Noldus FaceReader~\cite{den2005facereader};
\textbf{DVs:}
\textsf{PBI} = Privacy Behavioral Intention~\cite{yang2009influence}; 
\textsf{IUIPC} = Internet users' information privacy concerns~\cite{malhotra2004internet}.}
\label{fig:flowchart}
\end{figure*}

We developed the experiment design in a series of pretests, which established the validity and reliability of the overall design as well as the procedure and instruments used therein.
Figure~\ref{fig:flowchart} contains a flowchart depicting the development process in its entirety.

\subsection{Sampling}
Using flyers, and adverts on the crowdsourcing platforms- the participants involved in the user studies were recruited from different locations and at different time frames. The participants consist of institution's staff, students, workers on crowdsourcing platforms - Amazon Mechanical Turk and Prolific Academic. We adopted a simple sampling method based on participants' availability. The sample size were determined in an \textit{a priori} power analysis using \textsf{G*Power}.

\subsection{Ethics}

The studies reported here adhered to the institution's ethics guidelines and an ethics approval was obtained before the studies were conducted.

\paragraph{Affect Elicitation.}
The affect induction techniques used were not expected to adversely affect the participants' affect states. Stimulus video clips which have been tested in psychological studies and considered as standardized clips~\cite{coan2007handbook} were used. These clips were considered suitable for use in user studies which involve emotion~\cite{coan2007handbook}. The participants were not expected to experience any discomfort different from that encountered in daily life activities. At the end of the online user studies, the participants were provided with links to free online counseling services and advised to contact the researcher if they were agitated by contents of the user study.

\paragraph{Informed Consent and Opt-Out.}
During the recruitment process, participants were informed about the duration and requirements for the user studies. Participants were given an information sheet and a consent form which contained details about information that will be collected during the user studies. On the consent sheet they were presented with an opportunity to opt out at any stage without any penalties. All participants were given the opportunity to exercise informed consent. 

\paragraph{Deception.}
The true purpose of the user studies was not disclosed to the participants; rather they were informed that the aim of the user studies was to assess their opinions about online information management. They were presented with questions on privacy concerns, personality traits, privacy behavioral intentions and demographics. At the end of the user study, the real aim of the user study was explained during the debriefing session.

\paragraph{Compensation.}
All participants who completed either one or both parts of the user studies were duly compensated either with Amazon vouchers in person or cash using the provided payment platform.

\paragraph{Data Protection.}
We followed the institution's data protection policy. Participants personal information were anonymized and stored on an encrypted hard drive.

\subsection{Procedure}
We offer a comparison of the three studies conducted in Appendix~\ref{sec:studies}. The overall procedure  for the within-subjects experiment is as follows:

First the participants indicated their interest in a registration(pre task) form containing questions on privacy concerns, personality traits; The study was spread over two days;  in the first day, the participants carried out the following steps, i.e. 2--3. On the second day, the participants were first induced to a neutral state and then they completed steps 4 and 5. The reason for this was to minimize the carryover effects of the video stimuli and effect of questionnaire fatigue.

The procedure consists of the following steps, where Fig. 2 illustrates the key elements of the experiment design:
\begin{enumerate}
	\item Completion of pre-task questionnaire on demographics, alcohol/recreational drug use, IUIPC and CFIP surveys.
	\item Neutral state.
	\begin{enumerate}
		\item Induction of a neutral baseline affect state,
		\item DV questionnaires on privacy behavioral intentions,
		\item Manipulation check with PANAS-X,
		\item (Offline only) Manipulation check: Emotional Recognition (ER) and Facereader (FR) from video recording of the participant's face geometry.
	\end{enumerate}
	\item Affect State 1: Either happy or fear, determined by random assignment.
	\begin{enumerate}
		\item Show video stimulus to induce affect.
		\item  DV questionnaire on privacy behavioral intentions,
		\item	Manipulation check with PANAS-X.
		\item (Offline only) Manipulation check: Emotional Recognition (ER) and Facereader (FR) from video recording of the participant's face geometry.
	\end{enumerate}
	\item Affect State 2: Complement of Affect State 1.
	\begin{enumerate}
		\item Show video stimulus to induce affect.
		\item  DV questionnaire on privacy behavioral intentions,
		\item	Manipulation check with PANAS-X.
		\item (Offline only) Manipulation check: Emotional Recognition (ER) and Facereader (FR) from video recording of the participant's face geometry.
	\end{enumerate}
	\item a debriefing questionnaire, used to check for missed or misreported information, subjective thoughts during study session.
\end{enumerate}

\subsection{PBI Measurement}

We used the self-report PBI scale by Yang and Wang~\cite{yang2009influence} to measure the participants' privacy behavioral intentions. The reason for choosing this tool is because it considers privacy behavioral intention as a multi-dimensional construct, providing an all compassing assessment of PBI. This tool is made up of 14 questions which assess sub-scales: information disclosure intention (IDI), protection intention (PI) and transaction intention (TI). We have previously validated this tool in comparison with IUIPC~\cite{malhotra2004internet} considering dimensions of the Theory of Planned Behavior (TPB)~\cite{ajzen2011theory}. In our evaluation, we found that the privacy concern measured in IUIPC~\cite{malhotra2004internet} has characteristics of a long-term subjective norm. PBI on the other hand seemed more aligned with a short-term behavioral intention.

\subsection{Manipulation}
 In the three studies, all participants were required to watch standardized stimulus videos that induce neutral, fear and happy affect states. The three stimulus videos were selected from a list of stimulus videos recorded in the \textit{Handbook of emotion elicitation and assessment}~\cite{coan2007handbook}. As recommended in the Handbook, participants were asked to watch a scene from \textit{Alaska's Wild Denali}, to elicit a neutral affect state. To elicit happy, and fear affect states, participants were exposed to specified scenes from \textit{When Harry met Sally} and \textit{The Shining} respectively. These stimuli have been precisely defined and validated as standard measures to induce affect, as documented by Ray and Gross~\cite{ray2007emotion}.
 
\subsection{Manipulation Check}
We used a 15-item Positive and Negative Affect Schedule (PANAS-X) self-report questionnaire with a designated time horizon ``at this moment'') as main instrument to assess the manipulation success. 
For PANAS-X, we select the sub-scales \textsf{fear} and \textsf{joviality} to assess the affect states \textsf{fear} and \textsf{happiness}, respectively. According to Watson and Clark~\cite{watson1999panas}, the \textsf{joviality} scale is ``the longest and the most reliable of the lower order scales, with a median internal consistency estimate of $\alpha_{\mathsf{jov}} = .93$.'' The \textsf{fear} lower order scale is reported a median consistency of $\alpha_{\mathsf{fear}} = .87$.

While we also used psycho-physiological tools (Microsoft Emotional Recognition (ER) and Noldus FaceReader~\cite{den2005facereader}) in the lab, we report only the PANAS-X results in this paper as all the participants were assessed with this instrument.

%% file: sample.tex
\section{Sample}
\label{sec:sample}
\label{sec:studies}

The sample was combined from three studies with different properties but same methodology: \begin{inparaenum}[(i)]
\item Offline,
\item AMT, and
\item Prolific.
\end{inparaenum}

\paragraph{Study 1: Offline Lab study.}
The participants were recruited through flyers and emailing lists within the 
faculties of Social Sciences, Medical Sciences and Computer Science at the host university.
95 participants from Newcastle University registered online to participate in the study. Of those 95 participants, $N_L = 60$ participants completed the study by physically visiting the lab twice. In terms of ethnicity, 54\% of the participants were Caucasian, 26\% Asian and 20\% were African. In terms of classification of studies, 56.7\% of the participants were studying for a postgraduate degree, 37.7\% were studying for an undergraduate degree, 3.3\% of the participants had secondary school education and the remainder did not report their education background. Table~\ref{tab:descriptivesOffline} summarizes the descriptive statistics of this sub-sample.

\paragraph{Study 2: Online AMT study.}
The study was conducted in a series of sessions on Amazon Mechanical Turk. 
Out of 100 registrations, a total of 70 AMT workers completed both sessions of our study. 
However, 31 responses were found to be unsuitable, by which retained a sub-sample of $N_M = 39$ observations, described in Table~\ref{tab:descriptivesOnline}.

\paragraph{Study 3: Prolific Academic study.}
The same experiment  was conducted on Prolific with a considerably greater completion rate than in the AMT Study. In the first session of affect study we conducted on Prolific 50 submissions were made; out of which 39 completed the study. A second batch requesting for 200 participants was conducted. 217 completed the first part of the study; 211 returned and completed the study. 15 incorrectly completed surveys were excluded. 235 observations were included in this sub-sample, its descriptives being summarized in Table~\ref{tab:descriptivesProlific}. Table~\ref{prolific-demo} shows the demographics distribution in this sub-sample. 

\begin{table}[!htb]
	\caption{Demographics of Study 3 on Prolific}
	\label{prolific-demo}
	\begin{minipage}{.5\linewidth}
		\centering
		\begin{tabular}{@{}ll@{}}
			\toprule
			\multicolumn{2}{l}{Age} \\ \midrule
			18-23 & 46.4\% \\
			24-29 & 28.1\% \\
			30-35 & 12.8\% \\
			36-41 & 6.4\% \\
			42-47 & 2.9\% \\
			48-53 & 1.3\% \\
			54+ & 2.1\% \\ \bottomrule
		\end{tabular}
	\end{minipage}%
	\begin{minipage}{.5\linewidth}
		\centering
		\begin{tabular}{@{}ll@{}}
			\toprule
			\multicolumn{2}{l}{Gender} \\ \midrule
			Female & 39.1\% \\ 
			Male & 60.9\% \\ \bottomrule
		\end{tabular}
	\end{minipage}
\end{table}

%% file: discussion.tex
\section{Discussion}

\subsection{Incidental affect impacts PBI and Protection Intention.}
We found that both fear and happy affect states caused an increase of Privacy Behavioral Intention as well as the sub-construct Protection Intention. The magnitude of the effect of fear and happy states is roughly equal.

In the comparison between fear and happy affect states themselves, we did \emph{not} find a significant effect. More to the point, we found that the effects between neutral and the fear/happy were 20-fold the size of the effects between fear and happy condition.
Similarly, Coopamootoo~\cite{coopamootoo2017wip} and Fordyce et al~\cite{fordyce2018investigation} observed non-significant effects of happiness and fear on privacy behavioral intentions and password choice, respectively, without having provided a plausible explanation. 

Furthermore, we observed a cross-over interaction in the fear measurements, but not in the joviality measurements, shedding further light on the low effect between fear and happy conditions. 

\subsection{PBI sub-constructs are affected differently.}
While Protection Intention was significantly affected by fear as well as happy conditions, we found much smaller and, then, non-significant effects on Information Disclosure Intention and Transaction Intention. That these emotions act on the Protection Intention, but not on IDI or TI yields further evidence for the complex influence of incidental affect on PBI. These observations are again substantiated by observed interactions on sub-constructs. Musingly, we could say: ``It's complicated.''

The situation certainly calls for further investigation to ascertain how affect states impact security-relevant intentions and behaviors. From our analysis so far we conclude that simple comparisons of just two affects while ignoring the neutral state do not cut the mustard.

\subsection{Consulting the circumplex model for a hypothetical explanation: Arousal.}
We consider Russel's circumplex model of affect (cf. Figure~\ref{fig:circumplex},~\cite{russell1980circumplex}), acknowledging that it is not without contention~\cite{larsen1992promises}.
We find fear classified as negative-valence/high-arousal and happiness classified as positive-valence/high-arousal. Both emotions have the high arousal in common. We have not anticipated this effect before this study and can only offer a declared \emph{post-hoc} hypothesis: ``Arousal itself has a positive impact on Protection Intention and, hence, on Privacy Behavioral Intention.''

Is this hypothesis plausible?
Gro{\ss} et al.~\cite{GrCoAl2016} proposed Selye's arousal curve as an alternative explanation for the impact of cognitive effort and depletion on password choice they observed. Hence, arousal was considered a plausible explanation in affective-cognitive effects on security and privacy before. While Fordyce et al~\cite{fordyce2018investigation} explicitly investigated this problem by analyzing the impact of stress on password choice, the question of arousal itself was not settled. To the best of our knowledge, there has been no such investigation in online privacy, yet. However, given the results of this study we opine that arousal can no longer be ignored in similar studies and needs to be considered as a possible confounding variable.

%% file: limitations.tex
\subsection{Limitations}
\paragraph{Sampling.}
Our participants were recruited from different crowdsourcing platforms, AMT and Prolific Academic, as well as through flyers and e-mails for the lab experiment. 
The sampling method employed to recruit our participants was based on self-selection or availability, and not random.
The participants involved in our user studies were mainly from the US, possibly a norm based effect may have had an effect on the study results. 

\paragraph{PBI Instrument Properties.}
The questions for PBI sub-scales were not evenly distributed with two questions assessing information disclosure intentions, nine questions assessing protection intentions and four for transaction intentions~\cite{yang2009influence}. 

\paragraph{Mood Induction Properties.}
Though we used standard self reporting tools to measure affect states as well as standardized tools for mood induction, the mood induction could have been more robust and consistent if further affect induction techniques were employed to reinforce the induced affect states.

The standardized stimulus videos used were produced more than 10 years ago. As a result some participants were familiar with the film scenes and knew what to expect. This raises the question if an increased effect on affect states could have been observed if the films were not known.

Contrary to our expectations, we obtained reports from a small percentage of participants, who indicated that they enjoyed the fear stimulus film and reported high happiness scores after watching it. We imagine that this effect is rooted in the pop-culture co-notation of the films and personal preferences of users for certain genres.

%% file: conclusion.tex
\section{Conclusion}

We are first to offer an analysis of incidental affect on privacy behavioral intention (PBI) and its sub-constructs. Incidental affect refers to affect present in a user independent of the current task. Hence, our analysis is more general than earlier preliminary work on integral affect, that is, affect induced by the task at hand.

For the first time, our analysis contributes a systematic fine-grained comparison of neutral, happy and fear states of PBI and its sub-constructs as well as observed interactions. Our analysis offers a compelling explanation for low effect sizes observed between happy and fear states: Compared to neutral states, \emph{both} affect states cause an increased PBI as well as an increased Protection Intention of similar magnitude. Consequently, earlier analyses that only compared fear and happy states ended up with only registering tiny non-significant effects.

Our research raises further questions on a wider range user studies with mood/affect induction and binary comparisons between affect conditions, incl. work-in-progress papers on integral affect~\cite{coopamootoo2017wip} and contributions on the impact of fear on password choice~\cite{fordyce2018investigation}.

%% file: acknowledgment.tex
\section*{Acknowledgments}
This study benefited from Newcastle University's psycho-physiological eye tracking lab.
We are indebted to the Newcastle University's School of Computing for the offering the funds for the participant compensation.
Thomas Gro{\ss} was supported by the \CASCAde.